\documentstyle[12pt]{article}

\catcode`\@=11
\def\numberbysection{\@addtoreset{equation}{section}
        \def\theequation{\thesection.\arabic{equation}}}
\def\beq{\begin{equation}}
\def\eeq{\end{equation}}
\def\barr{\begin{eqnarray}}
\def\earr{\end{eqnarray}}
\def\dou{ 1 \: \! \! \! {\rm  I}}
\def\k{\kappa}
\def\kt {\widetilde\kappa}
\numberbysection
\begin{document}
\begin{titlepage}
\begin{center}
\hfill DFF 228/5/95\\
\vskip 1.in
{\large \bf Spontaneous Symmetry Breaking \\ in the Non-Abelian Anyon Fluid}
\vskip 0.5in
Andrea CAPPELLI and Paolo VALTANCOLI
\\[.2in]
{\em Dipartimento di Fisica, Universit\`a di Firenze\\
and INFN, Sezione di Firenze, \\
Largo E. Fermi, 2 \\
I - 50125 , Firenze ,Italy}
\end{center}
\vskip .5in
\begin{abstract}
We study the theory of non-relativistic matter coupled to the non-Abelian
$U(2)$ Chern-Simons gauge field in $(2+1)$ dimensions.
We adopt the mean field approximation in the current-algebra formulation
already applied to the Abelian anyons. We first show that this method is
able to describe both ``boson-based'' and ``fermion-based'' anyons and
yields consistent results over the whole range of fractional statistics.
In the non-Abelian theory, we find a superfluid (and superconductive) phase,
which is smoothly connected with the Abelian superfluid phase originally
discovered by Laughlin. The characteristic massless excitation is the
Goldstone particle of the specific mechanism of spontaneous symmetry breaking.
An additional massive mode is found by diagonalizing the non-local, non-Abelian
Hamiltonian in the radial gauge.
\end{abstract}
\vskip .3in
June 1995 \hfill\\
\end{titlepage}
\pagenumbering{arabic}
\section{Introduction}

The dynamics of anyons - particle with fractional statistics in $(2+1)$
dimensions \cite{wilczek} - has been considerably investigated in the past few
years. These collective excitations can arise in planar
condensed-matter systems like the fractional quantum Hall effect \cite{prange}
and the high-temperature superconductivity \cite{hi-tc}.
An effective field theory for anyons is obtained by coupling
non-relativistic matter particles - either bosons or fermions - to
the Abelian Chern-Simons gauge field, which provides the statistical
interaction \cite{jackiw} \cite{mich}.

The remarkable property of {\it superfluidity} is exhibited by anyons in the
thermodynamic limit at constant density \cite{fhl}:
the anyon {\it fluid} possesses
a ground state with uniform density and a massless longitudinal excitation.
This is a Goldstone mode which gives rise to superconductivity by the usual
Higgs mechanism when coupled to the physical electromagnetic field.
This theory was originally proposed by Laughlin \cite{fhl} for explaining
the  high-temperature superconductivity of cuprates \cite{hi-tc}. However,
the explicit breaking of P and T symmetries by
the fractional statistics \cite{wilczek} has not been confirmed by the
experiments so far \cite{laughrev}.

Independently of its physical application to high-temperature
superconductivity, we believe
that the anyon superfluid is very interesting and deserves a deeper analysis.
Few non-perturbative, semiclassical, ground states are known in field
theory, thus any new one is worth understanding for its own sake.
This may find wider applications than the original physical problem, as
it has occurred to spontaneous symmetry breaking.
Actually, the Anyon fluid is closely related to the usual superfluid,
because both exhibit the spontaneous breaking of the
$U(1)$ {\it global} symmetry related to particle number conservation.

In this paper, we show that the anyon superfluidity also
arises in the presence of a {\it non-Abelian} Chern-Simons interactions.
We consider the simplest case of particles having a {\it isospin}
$1/2 $ quantum number with $U(2)$ gauge symmetry. The application of
this theory to some $(2+1)$-dimensional physical
problems has been discussed in ref. \cite{froh}.
Here, we solve it in the mean field approximation, describe
the {\it specific} mechanism of spontaneous symmetry breaking and
study the low-energy excitations.

In section two, we review the mean field approximation \cite{fhl}
which allows to describe the non-perturbative ground state of the Abelian
anyons.
One assumes self-consistently that the matter density is spatially
uniform and obtains a uniform magnetic field by the Chern-Simons Gauss law,
$\ \langle \rho \rangle = - \langle B \rangle/\k\ $, where $\k\ $ is the
coupling constant. Thus, the particles uniformly fill up
the Landau levels determined by this mean magnetic field.
The quadratic fluctuations around the mean field can be
described \cite{carlo} by using the Dashen-Sharp current algebra formalism
\cite{sharp}; their diagonalization by a Bogoliubov
transformation produces a relativistic longitudinal excitation at low
energy, as in the familiar case of the superfluid \cite{flu}.
This is by far the simplest method for describing anyon superfluidity
\cite{fhl}.

While reviewing this method \cite{carlo}, we clarify one property
of the Chern-Simons interaction in the Hamiltonian formulation.
This long-range, topological interaction produces non-trivial boundary
effects, whose strength depends on the type of mean field ground state.
These boundary effects can be removed by normal-ordering the Hamiltonian,
but produce an effective {\it local} interaction, which is
ground-state dependent.
This property is crucial for describing both ``boson-based" and
``fermion-based" anyons within the current algebra approach.
Actually, the two descriptions of anyons are more accurate for fractional
statistics $\theta/\pi\sim 0$ and $\theta/\pi\sim 1$,
respectively, and agree at the mid-point of {\it semions}
$(\theta/\pi=1/2)$: their combination yields a consistent approximation
for all values of the statistics. In particular, we obtain the
approximate second-order ground-state energy.

In section three, we extend this method to the $U(2)$ non-Abelian Chern-Simons
interaction with two independent coupling constants, $\kappa_{U(1)} = \kappa$
and $\kappa_{SU(2)} = {\widetilde\kappa}$.
The mean field approximation produces two copies of Landau levels for
isospin-up and isospin-down matter, which have opposite contributions to the
mean iso-magnetic field.
For $\k\kt >0$, the ground-state configuration corresponds
 to equal populations of spin-up and spin-down particles, and to a vanishing
iso-magnetic field; if $1 / \kappa \rightarrow 0$, this ground state is P and
T invariant because parity-violating effects cancel between the two populations
- only excitations can break P and T explicitly. Models of this kind have been
discussed in refs. \cite{disass}.

Here we describe a different phase of the system, which exists for
$\kappa\kt  <0$ and $1/|\kt | < 4/|\kappa|$, and has a ground state with
maximally unbalanced populations. This phase is continuously connected to the
Abelian theory by tuning $1/|\kt |\to 0$.
The ground state breaks spontaneously the $U(2)$  global
symmetry to a $U(1)$ subgroup,
as in the Standard Model of electroweak interactions \cite{iz}.
We find it interesting that the low-energy dynamics of a non-Abelian gauge
theory can be solved in closed form in a toy model for spontaneous
symmetry breaking. Moreover, the dynamics of non-Abelian anyons has
not been much investigated so far \cite{jackiw}\cite{naa}.

In section four, we discuss the low-energy collective excitations above the
mean-field ground state. The quadratic expansion of the Hamiltonian, written
in terms of non-Abelian currents, consists of two independent parts,
corresponding to matter-density and isospin-density fluctuations, respectively.
The former fluctuations are massless and similar to those of the
Abelian anyon fluid.
The latter have a non-trivial, yet solvable, non-Abelian dynamics;
we solve explicitly the Gauss law constraint by using the radial gauge
$ \sum_{i=1}^2\ x^i A^a_i = 0 $, which maintains manifest rotation invariance
and breaks translation invariance \cite{rad}. The non-Abelian Hamiltonian
resembles a non-local deformation of the Landau-level Hamiltonian,
because isospin-flip excitations feel the mean
iso-magnetic field and, moreover, self-interact.
We obtain the complete spectrum and show that
this is gapful and discrete, with Gaussian fall-off of correlations.
Special care is paid to the gauge invariance of the set of physical states,
which is actually translational invariant.
This massive excitation does not spoil the Laughlin superconductivity
mechanism, because both low energy excitations become gapful upon coupling
to the physical electro-magnetic field.

In section five, the Abelian and
non-Abelian anyon superfluidities are explained in terms of the spontaneous
breaking of the global gauge symmetries, and the specific mechanisms are
compared with those of the Higgs and Standard Models of four
dimensional gauge theories \cite{iz}.
Finally, in the conclusion, we discuss
other possible physical applications of the non-Abelian anyon fluid.
In the appendix, we collect some additional
informations on the eigen-functions of the non-Abelian Hamiltonian.

\section{The mean field approximation in terms of currents}

\subsection{Hamiltonian and Abelian current algebra}

In this section, we review the mean field for the Abelian anyon fluid
in the current algebra approach of ref.\cite{carlo}.
The Lagrangian for non-relativistic matter
coupled to the Abelian Chern-Simons gauge field is \cite{jackiw},
\beq
{\cal L} = i \Psi^{\dagger} D_0 \Psi - \frac{1}{2m}
{(D_i \Psi )}^{\dagger} D_i \Psi
+ \frac{\kappa}{2} \epsilon^{\alpha\beta\gamma} A_{\alpha} \partial_{\beta}
A_{\gamma} \ ,
\label{lag} \eeq
where $ D_\mu = \partial_\mu + i A_\mu $ is the covariant derivative of the
gauge field
and $\Psi$ is the non-relativistic matter field\footnote{
We choose units $c=\hbar=1$ and set the electric charge equal to one.
We use the metric $\eta_{\mu\nu} = diag ( 1 , - 1 , - 1 ) $,
while for two-dimensional expressions we use $\delta_{ij}$,
i.e. $A_\mu = ( A_0 , - A^i = - A_i )$ and $x^\mu=(x^0, x^i=x_i)$.}.
The equation of motion for the matter field can be used to derive the
Hamiltonian,
\beq
{\cal H} = \int \ d^2{\bf x} \ \frac{1}{2m} {( D_i \Psi )}^{\dagger}
{( D_i \Psi )}  \ ,
\label{ham}\eeq
while the equations for the gauge field are,
\begin{eqnarray}
- F_{12} &=& B = \epsilon_{ij} \partial_i A_j = - \frac{1}{\k} \ \rho
\qquad ( { \rm Gauss \ \ law } ), \nonumber\\
F_{0i} &=& \partial_0 A_i + \partial_i A_0 = - \frac{1}{\k} \
\epsilon_{ij} J^j \ .
\label{gau} \end{eqnarray}
The conserved, gauge-invariant matter current
$J^\mu = ( \rho, J^i = J_i )$ is given by
\beq
\rho = \Psi^{\dagger} \Psi \ , \qquad\quad
J_i = \frac{1}{2im} ( \Psi^{\dagger} D_i \Psi - {( D_i \Psi )}^{\dagger}
\Psi )\ .
\label{mat} \eeq
The Chern-Simons field has no local physical degrees of freedom, thus
it can be solved in terms of the matter field at equal time by choosing
a complete gauge fixing.
The Coulomb gauge $\partial_i A_i = 0$ has been often used for non-relativistic
theories; here, we prefer the (spatial) radial gauge \cite{rad},
\beq
x^i A_i = 0 \ ,
\label{axi}\eeq
because it is better suited for the non-Abelian theory
discussed in the next section. Actually, any choice of gauge is equivalent
for the Abelian theory, because it will be described in terms of
gauge-invariant quantities.
The solutions of gauge field equations (\ref{gau}) in the radial gauge are
\cite{rad},
\begin{eqnarray}
A_0 & = & - \frac{1}{\k }\ \frac{1}{x \cdot \partial }\ \epsilon_{ij} x^i J_j
\ ,\nonumber \\
A_i & = & \frac{1}{\kappa}\ \frac{1}{1+ x \cdot \partial }\
\epsilon_{ij} x^j \rho \ ,
\label{inv} \end{eqnarray}
and do not actually involve time derivatives. A precise meaning of
the operator $ {( x \cdot \partial )}^{-1} $ is not important here,
and will be discussed in section four.
By using (\ref{inv}), we can write the Hamiltonian in terms of matter fields
only, and quantize it by requiring the {\it bosonic} commutation relations
\beq
\left[ \Psi ( {\bf x}, t ), \Psi^{\dagger} ( {\bf y}, t ) \right] =
\delta ( {\bf x} - {\bf y} ) \ .
\label{com} \eeq

Let us now choose the current $J_i$ and the density $\rho$ as
basic variables. Their algebra can be computed by using (\ref{com}) and
(\ref{inv}), and reads,
\begin{eqnarray}
 \left[ \rho ({\bf x}) , \rho ({\bf y}) \right] \ \ & = &  0  \ ,
\nonumber \\
 \left[ \rho ({\bf x}) , J_i ({\bf y}) \right] \ & = &
\frac{1}{im} {\partial\over \partial x^i} \left(
\delta ({\bf x} - {\bf y}) \rho ({\bf x}) \right)
\ ,\nonumber \\
\left[ J_i ({\bf x}) , J_j ({\bf y}) \right]  & = &  \frac{1}{im} \left[
{\partial\over\partial x^j}
\left( \delta ({\bf x} - {\bf y}) J_i ({\bf x}) \right) -
{\partial\over\partial y^i}
\left( \delta ({\bf x} - {\bf y}) J_j ({\bf y}) \right) \right]
\label{cr} \end{eqnarray}
An important property of this algebra is its {\it independence} of the
Chern-Simons coupling constant, which only appears in the
representation of the algebra (the states) and in the {\it normal-ordered}
Hamiltonian \cite{carlo}. The Hamiltonian (\ref{ham}) can also be written
in terms of currents as follows \cite{sharp},
\beq
{\cal H} = \int  d^2{\bf x} \ {1\over 8m}
\left( \partial_i \rho + 2im J_i \right)^{\dagger}\ \frac{1}{\rho}
\left( \partial_i \rho + 2im J_i \right) \ .
\label{cur} \eeq

\subsection{Mean field approximation}

Let us assume that the ground state $|\Omega \rangle$ has a spatially uniform
density,
\beq
\langle \Omega | \rho ({\bf x}) | \Omega \rangle = \rho_0 ,
\qquad \langle \Omega | J_i ({\bf x}) | \Omega \rangle = 0 \ ,
\label{vev} \eeq
which corresponds to a uniform magnetic field,
\beq
\langle \Omega | B({\bf x}) | \Omega \rangle = B_0 =
- \frac{1}{\kappa} \rho_0 \ ,
\label{mf}\eeq
by the Gauss law (\ref{gau}).
Next, we seek for self-consistency of this hypothesis in the approximate
quantum theory. By decoupling matter $(\rho)$ and field $(B)$ fluctuations
in (\ref{mf}), we quantize the field $\Psi$ in the external average magnetic
field $B_0$:
\beq
{\cal H} \rightarrow {\cal H}^{(0)} =
\int \ d^2 {\bf x} \ \frac{1}{2m} {( D^{(0)}_i \Psi )}^{\dagger}
(D^{(0)}_i \Psi ) \ ,
\label{lan} \eeq
where $ D^{(0)}_i = \partial_i - i A_i^{(0)} $, and  $ A_i^{(0)} =
\epsilon_{ij} x^j \ {\rho_0 / 2\k} $.
This is the well-known Hamiltonian of the Landau levels
in the so-called symmetric gauge \cite{ctz}. The one-particle energy
of the $n$-th Landau level is,
\beq
\epsilon_n = \frac{B_0}{m} ( n + \frac{1}{2} ) \ ,
\label{en}\eeq
and the eigen-functions of the lowest level are
\beq \psi_{0,\ell}({\bf x}) = \frac{1}{\lambda \sqrt{\pi \ell!}} \
\left(\frac{z}{\lambda}\right)^\ell  \ {\rm e}^{ - |z|^2 / 2\lambda^2 } \ ,
\qquad \left(z=x^1+ix^2 \right)\ ,
\label{eig}\eeq
where $ \lambda = \sqrt{2/{eB_0}} $ is the magnetic length, and $\ell$ is the
angular momentum. Note that these eigen-functions satisfy
\beq
( D^{(0)}_1 + i D^{(0)}_2 ) \psi_{0,\ell} = 0 \ .
\label{vac}\eeq
The angular momentum orbitals ( \ref{eig} ) have degenerate energy, and their
number is $(B_0 A/2\pi)$ in a finite domain of area $A$ (independent of $n$).

The mean field hypothesis is self-consistent for all the ground states
of ${\cal H}^{(0)}$ with $N$ particles which have  uniform density.
For bosonic matter, these have been found in ref.\cite{carlo},
and correspond to filling each Landau orbital of the lowest level
with the same number $n$ of particles.
Actually, using (\ref{eig},\ref{vac}), one can compute
\beq
\lim_{N \to \infty} \langle \Omega | \rho ( {\bf x}) | \Omega \rangle  =
\frac{n B_0}{2\pi} =\rho_0 \ ,\qquad
\lim_{N \to \infty} \langle \Omega | J_i ( {\bf x})  | \Omega \rangle = 0 \ .
\label{uni} \eeq
These values for $\rho_0$ agree with the Gauss law (\ref{gau}),
provided that $\kappa = - n/2\pi$.
Therefore, the mean field approximation is self-consistent for these
integer values of the coupling constant.
The ground-state energies, for a system of area $A$, are obtained from
(\ref{en}),
\beq
E_0^{(0)} = \epsilon_0\ N = \frac{\pi}{nm}\ \rho^2_0\ A \ ,\qquad
\left({\rm boson-based \ anyons }, \ \ \kappa = - \frac{n}{2\pi} \right)\ .
\label{gse} \eeq

The next order of the mean-field approximation is given by the
quadratic fluctuations of the density and the current. The Hamiltonian
(\ref{ham}) must be expanded quadratically and normal-ordered in the
thermodynamic limit $ N \to\infty $.
The latter limit involves some subtle boundary effects which actually
determine the strength of the effective {\it local} interaction
of fluctuations.
Actually, a more precise expression of (\ref{uni}) for large, but finite, $N$
can be obtained from (\ref{mat},\ref{eig}) \cite{ctz2},
\beq
\langle \Omega | \rho({\bf x}) | \Omega \rangle  \simeq
\rho_0 \ \Theta\left( {N\over \rho_0} - \pi|{\bf x}|^2 \right) \ ,\qquad
\langle \Omega | J_i ({\bf x}) | \Omega \rangle = -{1\over 2m}\
\epsilon_{ij}\ \partial_j \langle\Omega | \rho({\bf x}) | \Omega \rangle \ .
\label{uniN} \eeq
Namely, the density of a filled, finite, Landau level has the shape of
a droplet, with a chiral edge current, due to eq. (\ref{vac}). The contribution
of this edge current to the ground-state value of the Hamiltonian
in the form (\ref{cur}) is non-vanishing for $N\to\infty$,
and correctly gives the ground-state energy $E_0^{(0)}$ (\ref{gse}).
This boundary effect can be removed by rewriting the Hamiltonian.
Using an algebraic identity of the Bogomol'nyi type \cite{jackiw},
\beq
i \epsilon^{ij} {( D_i \Psi )}^{\dagger} D_j \Psi + m \epsilon^{ij}
\partial_i J^j + B \rho \equiv 0 \ ,
\label{bom} \eeq
the Hamiltonian (\ref{ham}) can be rewritten
\beq
{\cal H} = \int d^2 {\bf x}\ \frac{1}{2m}
\left[ {(D_i \Psi)}^{\dagger} (D_i \Psi) + i \alpha
\ \epsilon_{ij} {(D_i \Psi )}^{\dagger} D_j \Psi -
\frac{\alpha}{\kappa} \rho^2
\right]
\label{ord} \eeq
for any value of $\alpha$.
For $\alpha=1$, the derivative terms in (\ref{ord}) vanish on the ground
state, due to eq. (\ref{vac}); thus, there are no boundary effects.
The ground-state energy (\ref{gse}) is given by the local term $\rho^2$ only.
The new expression (\ref{ord}) of the Hamiltonian can be easily normal
ordered in the thermodynamic limit as follows:
\beq
: {\cal H} :\ \equiv {\cal H} - \langle\Omega |{\cal H}^{(0)}| \Omega\rangle
= {\cal H} - \frac{\pi}{nm} \int d^2 {\bf x} \rho^2_0 \ .
\label{sub} \eeq
Therefore, the Hamiltonian (\ref{ord}) with $\alpha = 1$ is adopted for
studying
the quadratic fluctuations. Note that the normal-ordering procedure has
produced an effective local interaction, which is the mean field approximation
of the long-range ``statistical repulsion'' of anyons. This statistical
repulsion
generates a positive energy density $E_0 > 0$ as in the case of free fermions.

Finally, we note that an attractive local interaction $(-g\rho^2/2)$
can also be included in the Hamiltonian for the anyon fluid (\ref{ham}),
(\ref{ord}) \cite{jackiw}\cite{dzf}.
The previous analysis can be extended for generic values of
$ g < 1/(m|\kappa|)$ .
At the ``self-dual'' point $ g = 1/(m|\kappa|) $, the local attraction
exactly balances the statistical repulsion, and there is phase transition
for the anyon fluid: non-trivial classical solutions with $E_0 = 0$
were found in ref.\cite{jackiw} and conformal invariance was shown to hold to
three-loop order in ref.\cite{dzf}.
The nature of the other phase $g>1/(m|\k|)$ is not presently understood.

\subsection{Quadratic fluctuations}

Following the approach of ref.\cite{carlo}, we study the quadratic
fluctuations using the variables $(\rho, J_i)$, satisfying the algebra
(\ref{cr}). Actually, we are only interested in representing this
algebra to leading order in the fluctuations, as well as expanding the
Hamiltonian (\ref{ord}) to quadratic order. To this effect, we
introduce a small parameter $\epsilon$ which keeps track of the size of
fluctuations,
\beq
\rho ({\bf x}) = \rho_0 + \epsilon {\hat \rho} ({\bf x})
+ O ( \epsilon^2 )\ , \
\ \ \ \ \ J({\bf x}) = \epsilon {\hat J} ({\bf x}) + O ( \epsilon^2 ) \ .
\label{flu} \eeq
By inserting this expansion in the second of the current commutators
(\ref{cr}), we obtain
\beq
\epsilon^2 \left[ {\hat \rho} ({\bf x}), {\hat J} ({\bf y}) \right] =
{\hbar} \frac{\rho_0}{im} \left( {\partial\over\partial x^i}
\delta ( {\bf x} - {\bf y} ) + O ( \epsilon ) \right) \ .
\label{app} \eeq
This shows that $\epsilon^2$ is of order $O(\hbar)$, so that we can neglect
the fluctuations in the r.h.s. of the commutators. The third commutator
in (\ref{cr}) can be similarly estimated:
\beq \epsilon^2 \left[ {\hat J}_i ({\bf x}), {\hat J}_j ({\bf y})
\right] = O ( \epsilon^3 ) \sim 0 \ .
\label{app2}\eeq
In general, the use of this $\epsilon$-expansion yields consistent
results for multiple commutators of the approximate algebra.

This approximate algebra, given by (\ref{app}), (\ref{app2}) and
$[\rho({\bf x}),\rho({\bf y})]=0$,
can be represented by a bosonic canonical field $\phi$, satisfying
$\left[ \phi ({\bf x}), \phi^{\dagger} ({\bf y})
\right] = \delta ( {\bf x} - {\bf y} ) $,
as follows,
\beq
{\hat\rho} ({\bf x}) = \sqrt{\rho_0} ( \phi + \phi^{\dagger} ) \ ,
\qquad {\hat{J}}_i ({\bf x}) =  \frac{\sqrt{\rho_0}}{2im} \partial_i
( \phi - \phi^{\dagger} ) \ .
\label{can} \eeq
(Note that the zero mode of $\phi$ is absent due to the
condition $ \int d^2 {\bf x} \ \rho ({\bf x}) = N $. )
The Hamiltonian (\ref{ord}), with $\ \alpha=1 $, can be written in terms of
currents, similarly to (\ref{cur}), and then expanded to quadratic order
in the fluctuations. The result is \cite{carlo}
\beq
{\cal H}^{(2)} = \int d^2 {\bf x} : \frac{1}{2m} \left[ \frac{1}{4\rho_0}
 {( {\hat K}_1 + i {\hat K}_2 )}^{\dagger} ( {\hat K}_1 + i {\hat K}_2 )
+ \frac{2\pi}{n}  {\hat\rho}^2 ({\bf x}) \right] :,   \label{qua} \eeq
where $ K_i \equiv \partial_i \rho + 2im J_i $.
By inserting the Fourier modes
\beq \phi ({\bf x}) = \int \frac{d^2 {\bf p}}{2\pi} e^{i {\bf p} \cdot {\bf x}}
\ a_{\bf p} , \ \ \ \ \ \left[ a_{\bf p} , a^{\dagger}_{\bf q}
\right] = \delta^2 ( {\bf p} - {\bf q} )
\label{fou} \eeq
one obtains,
\begin{eqnarray}
{\cal H}^{(2)} & = & \int d^2 {\bf p} :\left[ A_p a^{\dagger}_{\bf p}
a_{\bf p} + B ( a_{\bf p} a_{-{\bf p}} +
a^{\dagger}_{\bf p} a^{\dagger}_{-{\bf p}} ) \right]: \ , \nonumber \\
A_p & = &  \frac{p^2}{2m} + \frac{2\pi \rho_0}{nm}, \ \ \ \ \
B = \frac{\pi \rho_0}{nm} \label{dia}
\end{eqnarray}
The terms coming from the repulsive interaction can be normal-ordered
by the Bogoliubov transformation
\begin{eqnarray}
a_{\bf p} & = & \cosh \chi \ \alpha_{\bf p} + \sinh \chi \
\alpha^{\dagger}_{-{\bf p}} \ , \quad \chi=\chi(p)\ ,
\nonumber \\
a_{-{\bf p}}^{\dagger} & = & \sinh \chi \ \alpha_{\bf p} + \cosh \chi \
\alpha^{\dagger}_{-{\bf p}} \ .
\label{rot} \end{eqnarray}
This is an $SO(1,1)$ rotation in the $(a_{\bf p}, a^{\dagger}_{-{\bf p}})$
space which preserves the commutation relations.
The result is
\beq
{\cal H}^{(2)} = \Delta E_0^{(2)} + \int d^2 {\bf p} \ E_p \
\alpha_{\bf p}^{\dagger} \alpha_{\bf p} \ ,
\eeq
with
\beq
E_p = {| {\bf p} |\over m} \ \sqrt{ \frac{2\pi\rho_0}{n} +
\frac{p^2}{4} }\ \ \
{\buildrel p \rightarrow 0 \over \longrightarrow} \ \ v_s \ | {\bf p} | \ ,
\eeq
and
\beq
\Delta E_0^{(2)} = -{A\over(2\pi)^2}\ \int d^2 {\bf p} \ \frac{p^2}{4m}
\left( 1 + \frac{\eta}{p^2} - \sqrt{ 1 + \frac{2\eta}{p^2} }\ \right) \ ,
\qquad \eta = \frac{4\pi \rho_0}{n}
\label{deltae}\eeq
The quadratic fluctuations show a massless longitudinal excitation with sound
velocity \cite{carlo},
\beq
v_s = \frac{1}{m} \sqrt{ \frac{2\pi\rho_0}{n} } \ , \qquad
\left({\rm boson-based \ anyons}\ \ \kappa = - \frac{n}{2\pi}\ ,
\ {\theta\over\pi}  = \frac{1}{n}\right)
\label{velb} \eeq

The use of the Bogoliubov transformation \cite{carlo} makes manifest the
striking similarity between the anyon fluid and the usual superfluid
\cite{flu}. The latter is the canonical example for spontaneous breaking
of a global symmetry , the U(1) symmetry for particle number conservation.
Actually, the same spontaneous symmetry breaking occurs in the anyon fluid,
because the Bogoliubov rotated ground state, satisfying
$\alpha_{\bf p} |\widetilde\Omega\rangle =0$, does not have a well
defined particle number. The broken symmetry is the global $U(1)$ subgroup
of the gauge group\footnote{
Besides, it is the unique part of the gauge symmetry which can break [19].}.
Therefore, the anyon fluid gives an interesting new realization of the
Goldstone mechanism in non-relativistic field theory.
Note that the ``microscopic'' mechanism leading to
$\langle\rho\rangle= \rho_0$
is different from the Bose-Einstein condensation, and that there is no Higgs
phenomenon associated to the Chern-Simons field.
We shall discuss these differences in section five, together with the
results of the non-Abelian case.
The anyon fluid becomes a superconductor \cite{fhl} when is coupled to an
external electro-magnetic field, because the massless mode gives mass to the
photon by the usual Higgs mechanism.

\subsection{Fermion-based anyons}

It is interesting to extend these results to fermion-based anyons.
Suppose now that the matter field $\Psi$ satisfies canonical anti-commutation
relations. The mean field is again self-consistent (eq. (\ref{mf}))
for uniform fillings of the Landau levels, because each filled level
contributes a constant value to the density, away from the boundary
\cite{dunne}.
Due to Fermi statistics, we can put two spin-$1/2$ fermions per Landau
orbital, at most, and uniformly fill the lowest ${n}/{2}$ Landau levels,
where $n=2p+\sigma$, $\sigma=0,1$; if $n$ is odd ($\sigma=1$),
we fill the top level with one electron per orbital.
The resulting ground-state density is again given by (\ref{uni}),
and the allowed values of the Chern-Simons coupling
constant are $\k=-n/2\pi\ $, which correspond now to the fractional statistics
$\theta/\pi=1-1/n$.
The ground-state energy, obtained by (\ref{en}), is:
\barr
E^{(0)}_0 & =& {B_0 A\over 2\pi}\ \sum^{p - 1}_{k=0}
\left( 2\epsilon_k +\sigma\epsilon_p \right) = \left\{
\begin{array}{ll}
{A\rho_0^2\pi\over 2 m} \ ,& n \ {\rm even} ,\\
{A\rho_0^2\pi\over 2 m}\left(1+{1\over n^2}\right)\ , & n \ {\rm odd},
\end {array} \right.
\nonumber\\
&\ &\left({\rm \ fermion-based \ anyons}, \
\ \ \kappa = - \frac{n}{2\pi} \right) \ .
\label{fba} \earr
This ground-state energy oscillates between even and odd values of $n$ and
correctly reproduces the energy of the filled Fermi sea for
$\kappa\to\infty$.

The Hamiltonian must be normal-ordered differently from (\ref{sub}),
because the ground-state energy is higher for fermion-based anyons
than for boson-based ones. Again, we can dispose of the boundary
terms in the ground-state expectation value of $\cal H$ by choosing the
parameter $\alpha$ in eq. (\ref{ord}) which gives vanishing derivative terms.
This is found to be $ \alpha = n/2 \ +(\sigma/2n)\ $ by using some
equations similar to (\ref{vac}).
As a consequence, fermion-based anyons have an effective local repulsion
different from the boson-based ones.
The discussion of quadratic fluctuations is the same as in the previous
bosonic case, because the current algebra is independent of the statistics.
We obtain the Hamiltonian (\ref{dia}) with
\beq
A_p^F = \frac{p^2}{2m} + \frac{\pi\rho_0}{m}
\left(1+{\sigma\over n^2}\right) \ , \qquad
B^F = \frac{\pi\rho_0}{2m} \left(1+{\sigma\over n^2}\right) \ ,
\qquad \sigma=n \ {\rm mod}\ 2\ ,
\label{fer} \eeq
leading to a massless mode with sound velocity,
\beq
v_s^F = \left\{
\begin{array}{ll}
{1\over m}\ \sqrt{\rho_0\pi}\ , & \ n \ {\rm even}\ , \\
{1\over m}\ \sqrt{\rho_0\pi\left(1+{1\over n^2}\right) } \ ,
& \ n \ {\rm odd}\ ,
\end{array} \right.
\left({\rm fermion-based\ anyons}\ , \ \kappa= -{n\over {2\pi}}\ ,
\ {\theta\over\pi} = 1 - {1\over n} \right) \ .
\label{velf} \eeq

This value is different (always lower) than the result of ref.\cite{fhl}
for fermion-based anyons, because these authors considered spinless fermions.
In the free fermion limit, $ v_s \to v_F / {\sqrt{2}} $, where
$ v_F $ is the velocity of particle-hole excitations at the Fermi surface:
thus, the mean field approximation picks up one particular value of the
continuum of massless particle-hole excitations with velocities
$ 0 < v_s < v_F $. Note also that this approach gives the same result for
boson-based and fermion-based anyons at the common midpoint of semions, with
statistics $ \theta / \pi = 1 / 2 $, for both $E_0^{(0)}$ and $v_s$
(eqs. (\ref{gse},\ref{velb}) and (\ref{fba},\ref{velf}), respectively).
This shows that the current-algebra approach can describe both types of
anyon constructions, with greater accuracy
in the regions $ \theta / \pi \simeq 0 $ and
$ \theta / \pi \simeq 1 $, respectively, corresponding to
$ 1 / \k = 2 \pi / n \rightarrow 0 $ in both cases \cite{carlo}.

\subsection{Ground-state energy}

The best approximation for the ground-state energy
$E^{(0)}_0 + \Delta E^{(2)}_0$ is obtained by combining
the boson-based expression  for $0\le\theta/\pi\le 1/2$ and
the fermion-based one for
$1/2\le\theta/\pi\le 1$. As discussed in ref.\cite{carlo},
the second-order contribution $\Delta E^{(2)}_0$ (\ref{deltae}) is
ultraviolet divergent and must be regularized by allowing a {\it finite size}
to anyons, $a\equiv 1/\Lambda$, where $\Lambda$ is the momentum cut-off.
Anyons are collective excitations which naturally have a finite size;
however, this length does not appear in the effective Chern-Simons
Lagrangian and must be supplemented otherwise.
Possibly, it could be self-consistently determined in the exact
solution of this theory, which is, however, not known at present.
Within the mean-field approximation, anyons have the size given by
the magnetic length $O\left(1/\sqrt{B_0}\right)$, which is the
minimal localization of particles in the Landau levels.
Therefore, we have,
\beq
a={1\over\Lambda}={\delta\over\sqrt{B_0}}=\delta\
\sqrt{n\over 2\pi\rho_0}\ ,\qquad \delta=O(1)\ ,
\label{defa}\eeq
where $\delta$ is a proportionality constant.
This defines the cut-off $\Lambda$ for both the boson-based
$\left(\theta/\pi=1/n_B\right)$ and fermion-based
$\left(\theta/\pi=1-1/n_F\right)$ anyons,
\beq
\Lambda^2_B={2\pi\rho_0\over \delta_B^2 n_B}\ ,
\qquad \Lambda^2_F={2\pi\rho_0\over \delta_F^2 n_F}\ .
\label{cutofs} \eeq
The boson-based and fermion-based ground-state energies match at the semion
point $\theta/\pi=1/2$ for the natural choice $\delta_F=\delta_B=\delta$;
the parameter $\delta=O(1)$ is left free.

We can integrate $\Delta E^{(2)}_0$ (\ref{deltae}) with the respective
cut-offs,
use $\eta_B=4\pi\rho_0/n_B$ (respectively,
$\eta_F= 2\pi\rho_0\left( 1+\sigma/n_F^2\right)$), and obtain,
\barr
E^{(2)}_0&=& E^{(0)}_0+\Delta E^{(2)}_0 \nonumber\\
&=& \left\{
\begin{array}{ll}
{\pi\rho_0^2A\over m}\left[
{\theta\over \pi} -\left( {\theta\over\pi} \right)^2
F\left(\left( 2\delta^2\right)^{-1}\right) \right]\ ,&
\ 0\le {\theta\over\pi}={1\over n_B} \le {1\over 2} \ ,
\\
{\pi\rho_0^2A\over 2m}\left[
1+{\sigma\over n^2_F} -{1\over 2} \left(1+{\sigma\over n^2_F}\right)^2
F\left( \left(
n_F\delta^2 \left(1 +{\sigma\over n^2_F}\right)\right)^{-1} \right) \right]\ ,&
\ {1\over 2}\le {\theta\over\pi}=1-{1\over n_F} \le 1 \ ,
\end{array}  \right.
\label{egs2}\earr
where $\sigma=n_F \ {\rm mod}\ 2\ $, and
\beq
2\ F(y)=y^2 +2y- \left(y+1\right)\sqrt{y^2+2y} +
\log\left\vert y+1+\sqrt{y^2+2y}\right\vert\ .
\eeq

It is interesting to discuss the qualitative behavior of the ground-state
energy as a function of $\theta/\pi$. The quadratic correction is negative
definite, as it should, and vanishes at the end points $\theta/\pi=0,1$,
where the leading expressions $E^{(0)}_0$ already gives the exact result.
Near free bosons, $\theta/\pi\sim 0$, the ground-state energy is quadratic in
$\theta/\pi$: this is a natural second-order result for bosons
interacting with strength $O(\theta/\pi)$.
Near the fermionic end, $\theta/\pi =1-1/n_F \to 1$,
the oscillations $O(1/n_F^2)$ between even and odd $n_F$ values become
subleading, and
the ground-state energy approaches linearly the Fermi energy from below,
$E^{(2)}_0\simeq E^F_0 \left(1- \vert 1-\theta/\pi\vert/2\delta^2\right)$.
The shape of the ground-state energy as a function of statistics
has been much investigated in the quantum mechanics of a finite
number $N$ of anyons \cite{amel}.
A direct comparison of these results with (\ref{egs2}) is, however, difficult,
because the quantum-mechanical excited states form a continuum
in the large $N$ limit. We thus find that the field theoretic approach
gives the best result for this quantity.

\section{The $U(2)$ non-Abelian anyon fluid}

We now consider non-relativistic matter carrying an {\it isospin} $1/2$
representation of index $r = 1,2$, which interacts with an $U(1) \times SU(2)$
Chern-Simons gauge field $ {\cal A}_\mu = ( A_\mu, A_\mu^a ) $,
$a = 1, 2, 3$, with couplings
$\kappa_{U(1)} \equiv\k $ and $\kappa_{SU(2)} \equiv \kt $.
The Lagrangian reads
\beq
{\cal L} = i \Psi^\dagger ( D_0 \Psi ) - \frac{1}{2} {(D_i \Psi)}^{\dagger}
( D_i \Psi ) + \frac{\tilde{\kappa}}{2} \epsilon^{\alpha\beta\gamma} \left(
A_\alpha^a \partial_\beta A_\gamma^a - \frac{1}{3} \epsilon_{abc}
A^a_\alpha A^b_\beta A^c_\gamma \right) + \frac{\kappa}{2}
\epsilon^{\alpha\beta\gamma} A_\alpha \partial_\beta A_\gamma \ ,
\label{lna} \eeq
where the summation over the isospin index is implicit and
the covariant derivative is
\beq
D_\mu = \partial_\mu + i {\cal A}_\mu = ( \partial_\mu + i A_\mu )
\dou + i A_\mu^a \frac{\sigma^a}{2} \ ,
\label{usu} \eeq
and $\sigma^a$ are the Pauli matrices.
By proceeding as in the Abelian case, we find the Hamiltonian
\beq
{\cal H} = \int d^2 {\bf x}\ {1\over 2m}\  {(D_i \Psi)}^{\dagger}(D_i \Psi)\ .
\label{nham} \eeq
The equations of motion for the Abelian part of the gauge field are again
given by (\ref{gau}) and those of the $SU(2)$ part are,
\begin{eqnarray}
- F_{12}^a & \equiv &  B^a = \epsilon_{ij} \left( \partial_i A_j^a +
\frac{1}{2} \epsilon^{abc} A^b_i A^c_j  \right) = - \frac{1}{\kt}\
\rho^a \qquad ({\rm Gauss \ \ law}), \nonumber \\
F_{0i}^a & = & \partial_0 A_i^a + \partial_i A_0^a - \epsilon_{abc}
A_0^b A_i^c = - \frac{1}{\tilde{\kappa}} \epsilon_{ij} J_j^a \ ,
\label{ngau} \end{eqnarray}
where the $SU(2)$ isospin density and current are given by,
\begin{eqnarray}
\rho^a & = & {\Psi}^{\dagger}\ \frac{\sigma^a}{2}\ \Psi\ , \nonumber \\
J^a_i & = & \frac{1}{2im} \left( \Psi^{\dagger} \frac{\sigma^a}{2} D_i \Psi
- {(D_i \Psi)}^{\dagger} \frac{\sigma^a}{2} \Psi \right) \ .
\end{eqnarray}
The covariant conservation law is
\beq
\left( {\cal D}_\mu J^\mu \right)^a \equiv
\partial_\mu J^{\mu,a} - \epsilon^{abc} A_\mu^b J^{\mu,c} = 0 \ .
\eeq

\subsection{ U(2) mean field approximation}

Let us look for a self-consistent approximation of the ground state which
displays uniform densities of both matter and isospin,
\begin{eqnarray}
\langle \rho \rangle = \rho_0 , \ \ \ \ \langle J_i \rangle = 0 \nonumber \\
\langle \rho^a \rangle = \rho_0^a , \ \ \ \ \langle J_i^a \rangle = 0
\label{nmf} \end{eqnarray}
Given that local gauge invariance cannot break spontaneously \cite{eli}, a
nonvanishing mean isospin density $\rho_0^a$ is not rigorously true in the
exact theory. However, it is possible in the mean-field approximation,
where local gauge invariance is reduced to the global one.
Therefore, we shall argue that the mean field theory correctly describes the
breaking of the $U(2)$ {\it global} gauge symmetry down to a $U(1)$ subgroup.
Correspondingly, the isospin quantum number will no longer be conserved.

We can rotate the isospin axes so that the mean isospin density become
$ \langle \rho^a \rangle = \delta^a_3 \ \tilde{\rho}_0 $.
Constant densities imply constant magnetic and
iso-magnetic fields, eq. (\ref{ngau}) and
\beq
\langle \Omega | B^a | \Omega \rangle \equiv
\delta^a_3\  \widetilde{B}_0\ = - \frac{1}{\kt}
\ \delta^a_3 \ \tilde{\rho}_0 \ ,
\eeq
respectively, and the corresponding $A_\mu^{(0)}$ fields. The zeroth-order
mean-field Hamiltonian thus reads
\beq
{\cal H} \rightarrow {\cal H}^{(0)} = \int d^2 {\bf x}\ {1\over 2m}\
{( D_i^{(0)} \Psi )}^{\dagger} ( D^{(0)}_i \Psi ) \ ,
\eeq
where
\beq
D^{(0)}_i = \partial_i - i A^{(0)}_i \ ,\quad
A^{(0)}_i = \epsilon_{ij} \frac{x^j}{2} \left( \frac{\rho_0}{\kappa} \dou +
\frac{\widetilde{\rho}_0}{2\kt }\ \sigma^3 \right) \ .
\eeq

This Hamiltonian describes two copies of Landau levels, one for each
isospin orientation, with different values of the magnetic field
$ B_{+} = ( B_0 +\ \widetilde{B}_0/2 ) $ and
$ B_{-} = ( B_0 -\ \widetilde{B}_0/2) $, respectively.
The number of Landau orbitals in an area A is then
$ N^{\pm}_L = {| B_{\pm} | A}/{2\pi} $.
Following the same steps of the Abelian case, we test the consistency of the
mean field hypothesis by constructing uniform ground states for
these Landau level problems. For boson-based anyons, we
fill uniformly the pair of first Landau levels with $n_{+}$ isospin-up
and $n_{-}$ isospin-down particles per orbital, respectively.
The equations relating densities and magnetic fields are
\begin{eqnarray}
-\k B_0  =  \rho_0 & =& n_{+} \frac{N_L^{+}}{A} + n_{-} \frac{N_L^{-}}{A} =
n_{+} \frac{| B_0 + \frac{\widetilde{B}_0}{2} |}{2\pi} +
n_{-} \frac{| B_0 - \frac{\widetilde{B}_0}{2} |}{2\pi} \nonumber \\
- \kt \widetilde{B}_0 = \widetilde{\rho}_0 & =& \frac{1}{2} \left(
n_{+} \frac{| B_0 + \frac{\widetilde{B}_0}{2} |}{2\pi} -
n_{-} \frac{| B_0 - \frac{\widetilde{B}_0}{2} |}{2\pi} \right) \ .
\label{dme} \end{eqnarray}
These equations relate the unknown quantities
$ ( n_{+}, n_{-},\widetilde{\rho}_0 )$ to the external parameters \hfill\break
$(\rho_0, \k , \kt )$. In the Abelian case, the uniform filling was
only possible for certain values of $(\rho_0, \kappa )$. Here, instead,
there is a one-parameter freedom, which we fix by minimizing the ground-state
energy\footnote{
Note that gauge invariance requires $4\pi\kt$ to be an integer [4].}.
This can be expressed in terms of the known data as follows:
\begin{eqnarray}
\frac{E_0^{(0)}}{A} &=&
\frac{(B_0 + \frac{\widetilde{B}_0}{2})^2}{4\pi m} n_{+} +
\frac{(B_0 - \frac{\widetilde{B}_0}{2})^2}{4\pi m} n_{-} \nonumber\\
&=& \frac{1}{2m} \left[
\left| \frac{\rho_0}{\k} + \frac{\widetilde{\rho}_0}{2\kt} \right|
\left( \frac{\rho_0}{2} + \widetilde{\rho}_0 \right) +
\left| \frac{\rho_0}{\k} - \frac{\widetilde{\rho}_0}{2\kt} \right|
\left( \frac{\rho_0}{2} - \widetilde{\rho}_0 \right) \right] \ .
\label{min} \end{eqnarray}
We study this expression as a function of the unknown $\widetilde{\rho}_0$,
with range $ | \widetilde{\rho}_0 | \le \rho_0 / 2 $, and locate its minima.
Choosing for convenience $\k < 0$ and varying $\kt$, we find
three phases for the non-Abelian theory:

i) $\kt < 0$. The minimum of energy is found for
$\widetilde{\rho}_0 = 0$;

ii) $\kt > 0$ and $1/|\kt| > 4/ |\k|$. The minimum is found for
$\widetilde{\rho}_0 = \pm \rho_0\ 2 \tilde{\kappa} / \k $;

iii) $\kt > 0$ and $1/|\kt| <4/ |\k|$. The minimum is found for
$\widetilde{\rho}_0 = \pm \rho_0 / 2 $, with energy
\beq
E^{(0)}_0 \left\vert_{\widetilde{\rho}_0 =\pm\rho_0/2} \right. =
{ \rho_0^2 A\over 2m}\
\left( \frac{1}{|\k|} - \frac{1}{4|\kt|} \right) \ .
\label{nagse}\eeq
We see that the zero-th order mean field approximation manufactures
a classical potential with non-trivial minima, as in the standard
cases of spontaneously broken symmetry (see fig. 1).


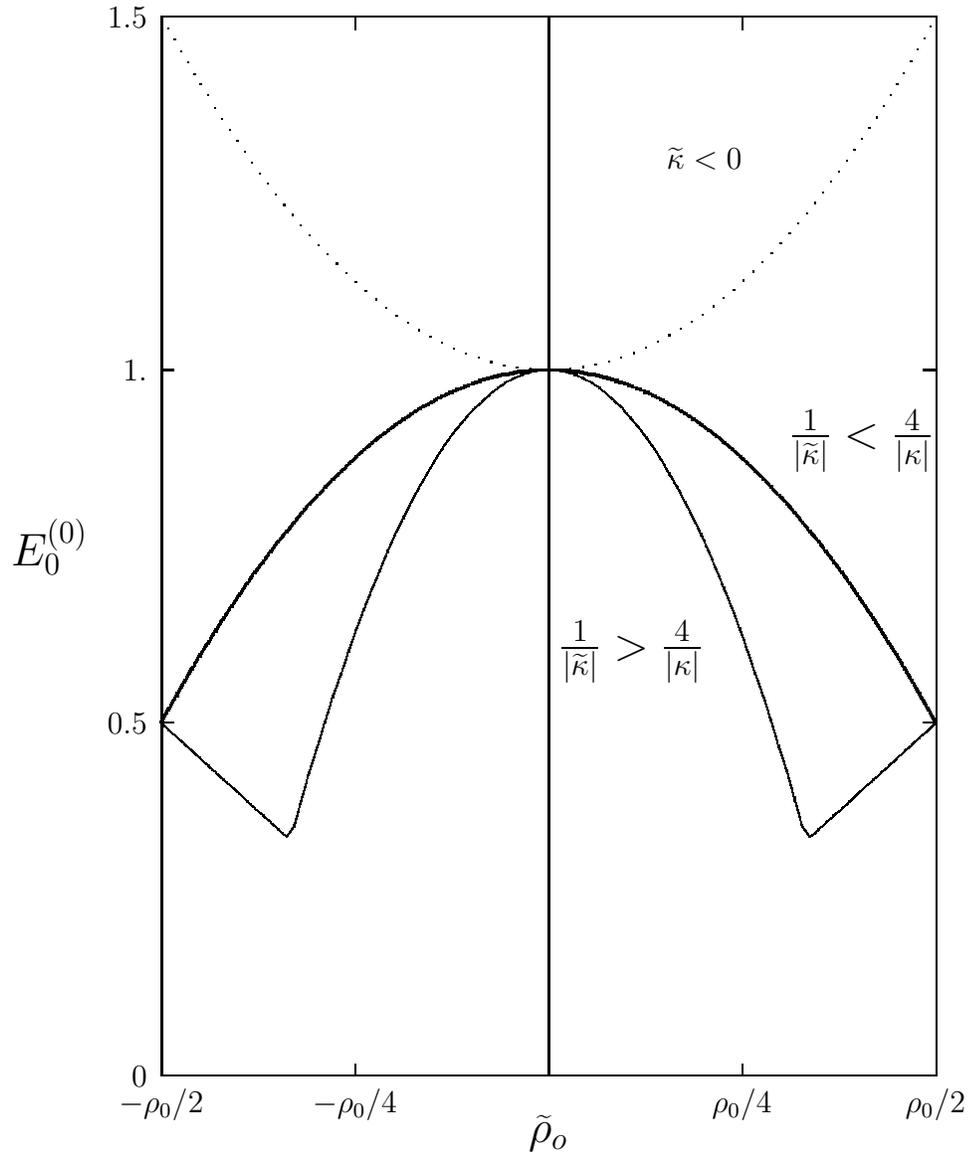
\begin{figure}
\setlength{\unitlength}{0.240900pt}
\ifx\plotpoint\undefined\newsavebox{\plotpoint}\fi
\sbox{\plotpoint}{\rule[-0.200pt]{0.400pt}{0.400pt}}%
\begin{picture}(1500,1800)(0,0)
\font\gnuplot=cmr10 at 10pt
\gnuplot
\sbox{\plotpoint}{\rule[-0.200pt]{0.400pt}{0.400pt}}%
\put(220.0,113.0){\rule[-0.200pt]{292.934pt}{0.400pt}}
\put(828.0,113.0){\rule[-0.200pt]{0.400pt}{400.858pt}}
\put(220.0,113.0){\rule[-0.200pt]{4.818pt}{0.400pt}}
\put(198,113){\makebox(0,0)[r]{$0$}}
\put(1416.0,113.0){\rule[-0.200pt]{4.818pt}{0.400pt}}
\put(220.0,668.0){\rule[-0.200pt]{4.818pt}{0.400pt}}
\put(198,668){\makebox(0,0)[r]{$0.5$}}
\put(1416.0,668.0){\rule[-0.200pt]{4.818pt}{0.400pt}}
\put(220.0,1222.0){\rule[-0.200pt]{4.818pt}{0.400pt}}
\put(198,1222){\makebox(0,0)[r]{$1.$}}
\put(1416.0,1222.0){\rule[-0.200pt]{4.818pt}{0.400pt}}
\put(220.0,1777.0){\rule[-0.200pt]{4.818pt}{0.400pt}}
\put(198,1777){\makebox(0,0)[r]{$1.5$}}
\put(1416.0,1777.0){\rule[-0.200pt]{4.818pt}{0.400pt}}
\put(220.0,113.0){\rule[-0.200pt]{0.400pt}{4.818pt}}
\put(220,68){\makebox(0,0){$\displaystyle -{\rho_0 /2}$}}
\put(220.0,1757.0){\rule[-0.200pt]{0.400pt}{4.818pt}}
\put(524.0,113.0){\rule[-0.200pt]{0.400pt}{4.818pt}}
\put(524,68){\makebox(0,0){$\displaystyle -{\rho_0 /4}$}}
\put(524.0,1757.0){\rule[-0.200pt]{0.400pt}{4.818pt}}
\put(1132.0,113.0){\rule[-0.200pt]{0.400pt}{4.818pt}}
\put(1132,68){\makebox(0,0){$\displaystyle {\rho_0 /4}$}}
\put(1132.0,1757.0){\rule[-0.200pt]{0.400pt}{4.818pt}}
\put(1436.0,113.0){\rule[-0.200pt]{0.400pt}{4.818pt}}
\put(1436,68){\makebox(0,0){$\displaystyle {\rho_0 / 2}$}}
\put(1436.0,1757.0){\rule[-0.200pt]{0.400pt}{4.818pt}}
\put(220.0,113.0){\rule[-0.200pt]{292.934pt}{0.400pt}}
\put(1436.0,113.0){\rule[-0.200pt]{0.400pt}{400.858pt}}
\put(220.0,1777.0){\rule[-0.200pt]{292.934pt}{0.400pt}}
\put(45,945){\makebox(0,0){{\Large$E_0^{(0)}$}}}
\put(828,23){\makebox(0,0){{\Large${\widetilde\rho_o}$}}}
\put(1132,1555){\makebox(0,0)[r]{$\widetilde\kappa <0$}}
\put(1193,1111){\makebox(0,0)[l]{
{\Large${1\over|\widetilde\kappa|}<{4\over|\kappa|}$}}}
\put(1071,779){\makebox(0,0)[r]{
{\Large${1\over|\widetilde\kappa|}>{4\over|\kappa|}$}}}
\put(220.0,113.0){\rule[-0.200pt]{0.400pt}{400.858pt}}
\put(220,668){\usebox{\plotpoint}}
\multiput(220.00,666.92)(0.496,-0.492){21}{\rule{0.500pt}{0.119pt}}
\multiput(220.00,667.17)(10.962,-12.000){2}{\rule{0.250pt}{0.400pt}}
\multiput(232.00,654.92)(0.590,-0.492){19}{\rule{0.573pt}{0.118pt}}
\multiput(232.00,655.17)(11.811,-11.000){2}{\rule{0.286pt}{0.400pt}}
\multiput(245.00,643.92)(0.543,-0.492){19}{\rule{0.536pt}{0.118pt}}
\multiput(245.00,644.17)(10.887,-11.000){2}{\rule{0.268pt}{0.400pt}}
\multiput(257.00,632.92)(0.543,-0.492){19}{\rule{0.536pt}{0.118pt}}
\multiput(257.00,633.17)(10.887,-11.000){2}{\rule{0.268pt}{0.400pt}}
\multiput(269.00,621.92)(0.543,-0.492){19}{\rule{0.536pt}{0.118pt}}
\multiput(269.00,622.17)(10.887,-11.000){2}{\rule{0.268pt}{0.400pt}}
\multiput(281.00,610.92)(0.539,-0.492){21}{\rule{0.533pt}{0.119pt}}
\multiput(281.00,611.17)(11.893,-12.000){2}{\rule{0.267pt}{0.400pt}}
\multiput(294.00,598.92)(0.543,-0.492){19}{\rule{0.536pt}{0.118pt}}
\multiput(294.00,599.17)(10.887,-11.000){2}{\rule{0.268pt}{0.400pt}}
\multiput(306.00,587.92)(0.543,-0.492){19}{\rule{0.536pt}{0.118pt}}
\multiput(306.00,588.17)(10.887,-11.000){2}{\rule{0.268pt}{0.400pt}}
\multiput(318.00,576.92)(0.590,-0.492){19}{\rule{0.573pt}{0.118pt}}
\multiput(318.00,577.17)(11.811,-11.000){2}{\rule{0.286pt}{0.400pt}}
\multiput(331.00,565.92)(0.543,-0.492){19}{\rule{0.536pt}{0.118pt}}
\multiput(331.00,566.17)(10.887,-11.000){2}{\rule{0.268pt}{0.400pt}}
\multiput(343.00,554.92)(0.496,-0.492){21}{\rule{0.500pt}{0.119pt}}
\multiput(343.00,555.17)(10.962,-12.000){2}{\rule{0.250pt}{0.400pt}}
\multiput(355.00,542.92)(0.543,-0.492){19}{\rule{0.536pt}{0.118pt}}
\multiput(355.00,543.17)(10.887,-11.000){2}{\rule{0.268pt}{0.400pt}}
\multiput(367.00,531.92)(0.590,-0.492){19}{\rule{0.573pt}{0.118pt}}
\multiput(367.00,532.17)(11.811,-11.000){2}{\rule{0.286pt}{0.400pt}}
\multiput(380.00,520.92)(0.543,-0.492){19}{\rule{0.536pt}{0.118pt}}
\multiput(380.00,521.17)(10.887,-11.000){2}{\rule{0.268pt}{0.400pt}}
\multiput(392.00,509.92)(0.543,-0.492){19}{\rule{0.536pt}{0.118pt}}
\multiput(392.00,510.17)(10.887,-11.000){2}{\rule{0.268pt}{0.400pt}}
\multiput(404.00,498.92)(0.539,-0.492){21}{\rule{0.533pt}{0.119pt}}
\multiput(404.00,499.17)(11.893,-12.000){2}{\rule{0.267pt}{0.400pt}}
\multiput(417.58,488.00)(0.492,0.712){21}{\rule{0.119pt}{0.667pt}}
\multiput(416.17,488.00)(12.000,15.616){2}{\rule{0.400pt}{0.333pt}}
\multiput(429.58,505.00)(0.492,1.832){21}{\rule{0.119pt}{1.533pt}}
\multiput(428.17,505.00)(12.000,39.817){2}{\rule{0.400pt}{0.767pt}}
\multiput(441.58,548.00)(0.492,1.832){21}{\rule{0.119pt}{1.533pt}}
\multiput(440.17,548.00)(12.000,39.817){2}{\rule{0.400pt}{0.767pt}}
\multiput(453.58,591.00)(0.493,1.567){23}{\rule{0.119pt}{1.331pt}}
\multiput(452.17,591.00)(13.000,37.238){2}{\rule{0.400pt}{0.665pt}}
\multiput(466.58,631.00)(0.492,1.703){21}{\rule{0.119pt}{1.433pt}}
\multiput(465.17,631.00)(12.000,37.025){2}{\rule{0.400pt}{0.717pt}}
\multiput(478.58,671.00)(0.492,1.616){21}{\rule{0.119pt}{1.367pt}}
\multiput(477.17,671.00)(12.000,35.163){2}{\rule{0.400pt}{0.683pt}}
\multiput(490.58,709.00)(0.493,1.408){23}{\rule{0.119pt}{1.208pt}}
\multiput(489.17,709.00)(13.000,33.493){2}{\rule{0.400pt}{0.604pt}}
\multiput(503.58,745.00)(0.492,1.530){21}{\rule{0.119pt}{1.300pt}}
\multiput(502.17,745.00)(12.000,33.302){2}{\rule{0.400pt}{0.650pt}}
\multiput(515.58,781.00)(0.492,1.444){21}{\rule{0.119pt}{1.233pt}}
\multiput(514.17,781.00)(12.000,31.440){2}{\rule{0.400pt}{0.617pt}}
\multiput(527.58,815.00)(0.492,1.358){21}{\rule{0.119pt}{1.167pt}}
\multiput(526.17,815.00)(12.000,29.579){2}{\rule{0.400pt}{0.583pt}}
\multiput(539.58,847.00)(0.493,1.250){23}{\rule{0.119pt}{1.085pt}}
\multiput(538.17,847.00)(13.000,29.749){2}{\rule{0.400pt}{0.542pt}}
\multiput(552.58,879.00)(0.492,1.229){21}{\rule{0.119pt}{1.067pt}}
\multiput(551.17,879.00)(12.000,26.786){2}{\rule{0.400pt}{0.533pt}}
\multiput(564.58,908.00)(0.492,1.229){21}{\rule{0.119pt}{1.067pt}}
\multiput(563.17,908.00)(12.000,26.786){2}{\rule{0.400pt}{0.533pt}}
\multiput(576.58,937.00)(0.492,1.142){21}{\rule{0.119pt}{1.000pt}}
\multiput(575.17,937.00)(12.000,24.924){2}{\rule{0.400pt}{0.500pt}}
\multiput(588.58,964.00)(0.493,1.012){23}{\rule{0.119pt}{0.900pt}}
\multiput(587.17,964.00)(13.000,24.132){2}{\rule{0.400pt}{0.450pt}}
\multiput(601.58,990.00)(0.492,1.013){21}{\rule{0.119pt}{0.900pt}}
\multiput(600.17,990.00)(12.000,22.132){2}{\rule{0.400pt}{0.450pt}}
\multiput(613.58,1014.00)(0.492,0.970){21}{\rule{0.119pt}{0.867pt}}
\multiput(612.17,1014.00)(12.000,21.201){2}{\rule{0.400pt}{0.433pt}}
\multiput(625.58,1037.00)(0.493,0.853){23}{\rule{0.119pt}{0.777pt}}
\multiput(624.17,1037.00)(13.000,20.387){2}{\rule{0.400pt}{0.388pt}}
\multiput(638.58,1059.00)(0.492,0.884){21}{\rule{0.119pt}{0.800pt}}
\multiput(637.17,1059.00)(12.000,19.340){2}{\rule{0.400pt}{0.400pt}}
\multiput(650.58,1080.00)(0.492,0.798){21}{\rule{0.119pt}{0.733pt}}
\multiput(649.17,1080.00)(12.000,17.478){2}{\rule{0.400pt}{0.367pt}}
\multiput(662.58,1099.00)(0.492,0.712){21}{\rule{0.119pt}{0.667pt}}
\multiput(661.17,1099.00)(12.000,15.616){2}{\rule{0.400pt}{0.333pt}}
\multiput(674.58,1116.00)(0.493,0.655){23}{\rule{0.119pt}{0.623pt}}
\multiput(673.17,1116.00)(13.000,15.707){2}{\rule{0.400pt}{0.312pt}}
\multiput(687.58,1133.00)(0.492,0.582){21}{\rule{0.119pt}{0.567pt}}
\multiput(686.17,1133.00)(12.000,12.824){2}{\rule{0.400pt}{0.283pt}}
\multiput(699.58,1147.00)(0.492,0.582){21}{\rule{0.119pt}{0.567pt}}
\multiput(698.17,1147.00)(12.000,12.824){2}{\rule{0.400pt}{0.283pt}}
\multiput(711.00,1161.58)(0.539,0.492){21}{\rule{0.533pt}{0.119pt}}
\multiput(711.00,1160.17)(11.893,12.000){2}{\rule{0.267pt}{0.400pt}}
\multiput(724.00,1173.58)(0.543,0.492){19}{\rule{0.536pt}{0.118pt}}
\multiput(724.00,1172.17)(10.887,11.000){2}{\rule{0.268pt}{0.400pt}}
\multiput(736.00,1184.58)(0.600,0.491){17}{\rule{0.580pt}{0.118pt}}
\multiput(736.00,1183.17)(10.796,10.000){2}{\rule{0.290pt}{0.400pt}}
\multiput(748.00,1194.59)(0.758,0.488){13}{\rule{0.700pt}{0.117pt}}
\multiput(748.00,1193.17)(10.547,8.000){2}{\rule{0.350pt}{0.400pt}}
\multiput(760.00,1202.59)(0.950,0.485){11}{\rule{0.843pt}{0.117pt}}
\multiput(760.00,1201.17)(11.251,7.000){2}{\rule{0.421pt}{0.400pt}}
\multiput(773.00,1209.59)(1.267,0.477){7}{\rule{1.060pt}{0.115pt}}
\multiput(773.00,1208.17)(9.800,5.000){2}{\rule{0.530pt}{0.400pt}}
\multiput(785.00,1214.60)(1.651,0.468){5}{\rule{1.300pt}{0.113pt}}
\multiput(785.00,1213.17)(9.302,4.000){2}{\rule{0.650pt}{0.400pt}}
\multiput(797.00,1218.61)(2.695,0.447){3}{\rule{1.833pt}{0.108pt}}
\multiput(797.00,1217.17)(9.195,3.000){2}{\rule{0.917pt}{0.400pt}}
\put(810,1220.67){\rule{2.891pt}{0.400pt}}
\multiput(810.00,1220.17)(6.000,1.000){2}{\rule{1.445pt}{0.400pt}}
\put(834,1220.67){\rule{2.891pt}{0.400pt}}
\multiput(834.00,1221.17)(6.000,-1.000){2}{\rule{1.445pt}{0.400pt}}
\multiput(846.00,1219.95)(2.695,-0.447){3}{\rule{1.833pt}{0.108pt}}
\multiput(846.00,1220.17)(9.195,-3.000){2}{\rule{0.917pt}{0.400pt}}
\multiput(859.00,1216.94)(1.651,-0.468){5}{\rule{1.300pt}{0.113pt}}
\multiput(859.00,1217.17)(9.302,-4.000){2}{\rule{0.650pt}{0.400pt}}
\multiput(871.00,1212.93)(1.267,-0.477){7}{\rule{1.060pt}{0.115pt}}
\multiput(871.00,1213.17)(9.800,-5.000){2}{\rule{0.530pt}{0.400pt}}
\multiput(883.00,1207.93)(0.950,-0.485){11}{\rule{0.843pt}{0.117pt}}
\multiput(883.00,1208.17)(11.251,-7.000){2}{\rule{0.421pt}{0.400pt}}
\multiput(896.00,1200.93)(0.758,-0.488){13}{\rule{0.700pt}{0.117pt}}
\multiput(896.00,1201.17)(10.547,-8.000){2}{\rule{0.350pt}{0.400pt}}
\multiput(908.00,1192.92)(0.600,-0.491){17}{\rule{0.580pt}{0.118pt}}
\multiput(908.00,1193.17)(10.796,-10.000){2}{\rule{0.290pt}{0.400pt}}
\multiput(920.00,1182.92)(0.543,-0.492){19}{\rule{0.536pt}{0.118pt}}
\multiput(920.00,1183.17)(10.887,-11.000){2}{\rule{0.268pt}{0.400pt}}
\multiput(932.00,1171.92)(0.539,-0.492){21}{\rule{0.533pt}{0.119pt}}
\multiput(932.00,1172.17)(11.893,-12.000){2}{\rule{0.267pt}{0.400pt}}
\multiput(945.58,1158.65)(0.492,-0.582){21}{\rule{0.119pt}{0.567pt}}
\multiput(944.17,1159.82)(12.000,-12.824){2}{\rule{0.400pt}{0.283pt}}
\multiput(957.58,1144.65)(0.492,-0.582){21}{\rule{0.119pt}{0.567pt}}
\multiput(956.17,1145.82)(12.000,-12.824){2}{\rule{0.400pt}{0.283pt}}
\multiput(969.58,1130.41)(0.493,-0.655){23}{\rule{0.119pt}{0.623pt}}
\multiput(968.17,1131.71)(13.000,-15.707){2}{\rule{0.400pt}{0.312pt}}
\multiput(982.58,1113.23)(0.492,-0.712){21}{\rule{0.119pt}{0.667pt}}
\multiput(981.17,1114.62)(12.000,-15.616){2}{\rule{0.400pt}{0.333pt}}
\multiput(994.58,1095.96)(0.492,-0.798){21}{\rule{0.119pt}{0.733pt}}
\multiput(993.17,1097.48)(12.000,-17.478){2}{\rule{0.400pt}{0.367pt}}
\multiput(1006.58,1076.68)(0.492,-0.884){21}{\rule{0.119pt}{0.800pt}}
\multiput(1005.17,1078.34)(12.000,-19.340){2}{\rule{0.400pt}{0.400pt}}
\multiput(1018.58,1055.77)(0.493,-0.853){23}{\rule{0.119pt}{0.777pt}}
\multiput(1017.17,1057.39)(13.000,-20.387){2}{\rule{0.400pt}{0.388pt}}
\multiput(1031.58,1033.40)(0.492,-0.970){21}{\rule{0.119pt}{0.867pt}}
\multiput(1030.17,1035.20)(12.000,-21.201){2}{\rule{0.400pt}{0.433pt}}
\multiput(1043.58,1010.26)(0.492,-1.013){21}{\rule{0.119pt}{0.900pt}}
\multiput(1042.17,1012.13)(12.000,-22.132){2}{\rule{0.400pt}{0.450pt}}
\multiput(1055.58,986.26)(0.493,-1.012){23}{\rule{0.119pt}{0.900pt}}
\multiput(1054.17,988.13)(13.000,-24.132){2}{\rule{0.400pt}{0.450pt}}
\multiput(1068.58,959.85)(0.492,-1.142){21}{\rule{0.119pt}{1.000pt}}
\multiput(1067.17,961.92)(12.000,-24.924){2}{\rule{0.400pt}{0.500pt}}
\multiput(1080.58,932.57)(0.492,-1.229){21}{\rule{0.119pt}{1.067pt}}
\multiput(1079.17,934.79)(12.000,-26.786){2}{\rule{0.400pt}{0.533pt}}
\multiput(1092.58,903.57)(0.492,-1.229){21}{\rule{0.119pt}{1.067pt}}
\multiput(1091.17,905.79)(12.000,-26.786){2}{\rule{0.400pt}{0.533pt}}
\multiput(1104.58,874.50)(0.493,-1.250){23}{\rule{0.119pt}{1.085pt}}
\multiput(1103.17,876.75)(13.000,-29.749){2}{\rule{0.400pt}{0.542pt}}
\multiput(1117.58,842.16)(0.492,-1.358){21}{\rule{0.119pt}{1.167pt}}
\multiput(1116.17,844.58)(12.000,-29.579){2}{\rule{0.400pt}{0.583pt}}
\multiput(1129.58,809.88)(0.492,-1.444){21}{\rule{0.119pt}{1.233pt}}
\multiput(1128.17,812.44)(12.000,-31.440){2}{\rule{0.400pt}{0.617pt}}
\multiput(1141.58,775.60)(0.492,-1.530){21}{\rule{0.119pt}{1.300pt}}
\multiput(1140.17,778.30)(12.000,-33.302){2}{\rule{0.400pt}{0.650pt}}
\multiput(1153.58,739.99)(0.493,-1.408){23}{\rule{0.119pt}{1.208pt}}
\multiput(1152.17,742.49)(13.000,-33.493){2}{\rule{0.400pt}{0.604pt}}
\multiput(1166.58,703.33)(0.492,-1.616){21}{\rule{0.119pt}{1.367pt}}
\multiput(1165.17,706.16)(12.000,-35.163){2}{\rule{0.400pt}{0.683pt}}
\multiput(1178.58,665.05)(0.492,-1.703){21}{\rule{0.119pt}{1.433pt}}
\multiput(1177.17,668.03)(12.000,-37.025){2}{\rule{0.400pt}{0.717pt}}
\multiput(1190.58,625.48)(0.493,-1.567){23}{\rule{0.119pt}{1.331pt}}
\multiput(1189.17,628.24)(13.000,-37.238){2}{\rule{0.400pt}{0.665pt}}
\multiput(1203.58,584.63)(0.492,-1.832){21}{\rule{0.119pt}{1.533pt}}
\multiput(1202.17,587.82)(12.000,-39.817){2}{\rule{0.400pt}{0.767pt}}
\multiput(1215.58,541.63)(0.492,-1.832){21}{\rule{0.119pt}{1.533pt}}
\multiput(1214.17,544.82)(12.000,-39.817){2}{\rule{0.400pt}{0.767pt}}
\multiput(1227.58,502.23)(0.492,-0.712){21}{\rule{0.119pt}{0.667pt}}
\multiput(1226.17,503.62)(12.000,-15.616){2}{\rule{0.400pt}{0.333pt}}
\multiput(1239.00,488.58)(0.539,0.492){21}{\rule{0.533pt}{0.119pt}}
\multiput(1239.00,487.17)(11.893,12.000){2}{\rule{0.267pt}{0.400pt}}
\multiput(1252.00,500.58)(0.543,0.492){19}{\rule{0.536pt}{0.118pt}}
\multiput(1252.00,499.17)(10.887,11.000){2}{\rule{0.268pt}{0.400pt}}
\multiput(1264.00,511.58)(0.543,0.492){19}{\rule{0.536pt}{0.118pt}}
\multiput(1264.00,510.17)(10.887,11.000){2}{\rule{0.268pt}{0.400pt}}
\multiput(1276.00,522.58)(0.590,0.492){19}{\rule{0.573pt}{0.118pt}}
\multiput(1276.00,521.17)(11.811,11.000){2}{\rule{0.286pt}{0.400pt}}
\multiput(1289.00,533.58)(0.543,0.492){19}{\rule{0.536pt}{0.118pt}}
\multiput(1289.00,532.17)(10.887,11.000){2}{\rule{0.268pt}{0.400pt}}
\multiput(1301.00,544.58)(0.496,0.492){21}{\rule{0.500pt}{0.119pt}}
\multiput(1301.00,543.17)(10.962,12.000){2}{\rule{0.250pt}{0.400pt}}
\multiput(1313.00,556.58)(0.543,0.492){19}{\rule{0.536pt}{0.118pt}}
\multiput(1313.00,555.17)(10.887,11.000){2}{\rule{0.268pt}{0.400pt}}
\multiput(1325.00,567.58)(0.590,0.492){19}{\rule{0.573pt}{0.118pt}}
\multiput(1325.00,566.17)(11.811,11.000){2}{\rule{0.286pt}{0.400pt}}
\multiput(1338.00,578.58)(0.543,0.492){19}{\rule{0.536pt}{0.118pt}}
\multiput(1338.00,577.17)(10.887,11.000){2}{\rule{0.268pt}{0.400pt}}
\multiput(1350.00,589.58)(0.543,0.492){19}{\rule{0.536pt}{0.118pt}}
\multiput(1350.00,588.17)(10.887,11.000){2}{\rule{0.268pt}{0.400pt}}
\multiput(1362.00,600.58)(0.539,0.492){21}{\rule{0.533pt}{0.119pt}}
\multiput(1362.00,599.17)(11.893,12.000){2}{\rule{0.267pt}{0.400pt}}
\multiput(1375.00,612.58)(0.543,0.492){19}{\rule{0.536pt}{0.118pt}}
\multiput(1375.00,611.17)(10.887,11.000){2}{\rule{0.268pt}{0.400pt}}
\multiput(1387.00,623.58)(0.543,0.492){19}{\rule{0.536pt}{0.118pt}}
\multiput(1387.00,622.17)(10.887,11.000){2}{\rule{0.268pt}{0.400pt}}
\multiput(1399.00,634.58)(0.543,0.492){19}{\rule{0.536pt}{0.118pt}}
\multiput(1399.00,633.17)(10.887,11.000){2}{\rule{0.268pt}{0.400pt}}
\multiput(1411.00,645.58)(0.590,0.492){19}{\rule{0.573pt}{0.118pt}}
\multiput(1411.00,644.17)(11.811,11.000){2}{\rule{0.286pt}{0.400pt}}
\multiput(1424.00,656.58)(0.496,0.492){21}{\rule{0.500pt}{0.119pt}}
\multiput(1424.00,655.17)(10.962,12.000){2}{\rule{0.250pt}{0.400pt}}
\put(822.0,1222.0){\rule[-0.200pt]{2.891pt}{0.400pt}}
\put(220,1777){\usebox{\plotpoint}}
\multiput(220,1777)(9.939,-18.221){2}{\usebox{\plotpoint}}
\put(240.37,1740.84){\usebox{\plotpoint}}
\put(250.78,1722.88){\usebox{\plotpoint}}
\put(261.08,1704.86){\usebox{\plotpoint}}
\put(271.46,1686.89){\usebox{\plotpoint}}
\multiput(281,1671)(11.312,-17.402){2}{\usebox{\plotpoint}}
\put(304.23,1633.95){\usebox{\plotpoint}}
\put(315.24,1616.36){\usebox{\plotpoint}}
\put(327.13,1599.36){\usebox{\plotpoint}}
\put(338.85,1582.23){\usebox{\plotpoint}}
\put(350.36,1564.96){\usebox{\plotpoint}}
\put(362.15,1547.88){\usebox{\plotpoint}}
\put(374.49,1531.20){\usebox{\plotpoint}}
\put(387.02,1514.65){\usebox{\plotpoint}}
\put(399.47,1498.04){\usebox{\plotpoint}}
\put(412.33,1481.75){\usebox{\plotpoint}}
\put(425.34,1465.58){\usebox{\plotpoint}}
\put(438.69,1449.70){\usebox{\plotpoint}}
\put(452.20,1433.94){\usebox{\plotpoint}}
\multiput(453,1433)(14.123,-15.209){0}{\usebox{\plotpoint}}
\put(466.28,1418.69){\usebox{\plotpoint}}
\put(480.46,1403.54){\usebox{\plotpoint}}
\put(495.14,1388.86){\usebox{\plotpoint}}
\put(510.10,1374.49){\usebox{\plotpoint}}
\put(524.98,1360.02){\usebox{\plotpoint}}
\multiput(527,1358)(15.300,-14.025){0}{\usebox{\plotpoint}}
\put(540.28,1346.01){\usebox{\plotpoint}}
\put(556.59,1333.18){\usebox{\plotpoint}}
\put(572.53,1319.89){\usebox{\plotpoint}}
\multiput(576,1317)(16.604,-12.453){0}{\usebox{\plotpoint}}
\put(589.06,1307.35){\usebox{\plotpoint}}
\put(606.60,1296.26){\usebox{\plotpoint}}
\put(623.87,1284.75){\usebox{\plotpoint}}
\multiput(625,1284)(18.275,-9.840){0}{\usebox{\plotpoint}}
\put(642.00,1274.66){\usebox{\plotpoint}}
\put(660.29,1264.86){\usebox{\plotpoint}}
\multiput(662,1264)(18.564,-9.282){0}{\usebox{\plotpoint}}
\put(678.92,1255.73){\usebox{\plotpoint}}
\put(697.95,1247.44){\usebox{\plotpoint}}
\multiput(699,1247)(19.690,-6.563){0}{\usebox{\plotpoint}}
\put(717.66,1240.95){\usebox{\plotpoint}}
\multiput(724,1239)(19.690,-6.563){0}{\usebox{\plotpoint}}
\put(737.43,1234.64){\usebox{\plotpoint}}
\put(757.56,1229.61){\usebox{\plotpoint}}
\multiput(760,1229)(20.514,-3.156){0}{\usebox{\plotpoint}}
\put(778.02,1226.16){\usebox{\plotpoint}}
\multiput(785,1225)(20.684,-1.724){0}{\usebox{\plotpoint}}
\put(798.63,1223.87){\usebox{\plotpoint}}
\put(819.32,1222.22){\usebox{\plotpoint}}
\multiput(822,1222)(20.756,0.000){0}{\usebox{\plotpoint}}
\put(840.05,1222.50){\usebox{\plotpoint}}
\multiput(846,1223)(20.694,1.592){0}{\usebox{\plotpoint}}
\put(860.74,1224.14){\usebox{\plotpoint}}
\put(881.32,1226.72){\usebox{\plotpoint}}
\multiput(883,1227)(20.514,3.156){0}{\usebox{\plotpoint}}
\put(901.72,1230.43){\usebox{\plotpoint}}
\multiput(908,1232)(20.136,5.034){0}{\usebox{\plotpoint}}
\put(921.81,1235.60){\usebox{\plotpoint}}
\put(941.58,1241.95){\usebox{\plotpoint}}
\multiput(945,1243)(19.690,6.563){0}{\usebox{\plotpoint}}
\put(961.18,1248.74){\usebox{\plotpoint}}
\put(980.15,1257.15){\usebox{\plotpoint}}
\multiput(982,1258)(18.564,9.282){0}{\usebox{\plotpoint}}
\put(998.74,1266.37){\usebox{\plotpoint}}
\put(1016.92,1276.37){\usebox{\plotpoint}}
\multiput(1018,1277)(18.275,9.840){0}{\usebox{\plotpoint}}
\put(1034.94,1286.63){\usebox{\plotpoint}}
\put(1052.21,1298.14){\usebox{\plotpoint}}
\multiput(1055,1300)(17.677,10.878){0}{\usebox{\plotpoint}}
\put(1069.71,1309.28){\usebox{\plotpoint}}
\put(1086.07,1322.06){\usebox{\plotpoint}}
\put(1102.01,1335.34){\usebox{\plotpoint}}
\multiput(1104,1337)(16.451,12.655){0}{\usebox{\plotpoint}}
\put(1118.30,1348.19){\usebox{\plotpoint}}
\put(1133.41,1362.41){\usebox{\plotpoint}}
\put(1148.39,1376.78){\usebox{\plotpoint}}
\put(1163.26,1391.26){\usebox{\plotpoint}}
\put(1177.93,1405.93){\usebox{\plotpoint}}
\multiput(1178,1406)(14.078,15.251){0}{\usebox{\plotpoint}}
\put(1192.02,1421.17){\usebox{\plotpoint}}
\put(1206.01,1436.51){\usebox{\plotpoint}}
\put(1219.51,1452.26){\usebox{\plotpoint}}
\put(1232.78,1468.22){\usebox{\plotpoint}}
\put(1245.81,1484.38){\usebox{\plotpoint}}
\put(1258.56,1500.75){\usebox{\plotpoint}}
\put(1271.02,1517.35){\usebox{\plotpoint}}
\put(1283.56,1533.89){\usebox{\plotpoint}}
\put(1295.81,1550.64){\usebox{\plotpoint}}
\put(1307.52,1567.78){\usebox{\plotpoint}}
\put(1319.03,1585.05){\usebox{\plotpoint}}
\put(1330.85,1602.10){\usebox{\plotpoint}}
\put(1342.56,1619.22){\usebox{\plotpoint}}
\put(1353.51,1636.86){\usebox{\plotpoint}}
\put(1364.32,1654.57){\usebox{\plotpoint}}
\multiput(1375,1671)(10.679,17.798){2}{\usebox{\plotpoint}}
\put(1396.60,1707.80){\usebox{\plotpoint}}
\put(1406.90,1725.82){\usebox{\plotpoint}}
\put(1417.35,1743.75){\usebox{\plotpoint}}
\put(1427.68,1761.75){\usebox{\plotpoint}}
\put(1436,1777){\usebox{\plotpoint}}
\sbox{\plotpoint}{\rule[-0.400pt]{0.800pt}{0.800pt}}%
\put(220,668){\usebox{\plotpoint}}
\multiput(221.41,668.00)(0.511,0.943){17}{\rule{0.123pt}{1.667pt}}
\multiput(218.34,668.00)(12.000,18.541){2}{\rule{0.800pt}{0.833pt}}
\multiput(233.41,690.00)(0.509,0.864){19}{\rule{0.123pt}{1.554pt}}
\multiput(230.34,690.00)(13.000,18.775){2}{\rule{0.800pt}{0.777pt}}
\multiput(246.41,712.00)(0.511,0.897){17}{\rule{0.123pt}{1.600pt}}
\multiput(243.34,712.00)(12.000,17.679){2}{\rule{0.800pt}{0.800pt}}
\multiput(258.41,733.00)(0.511,0.897){17}{\rule{0.123pt}{1.600pt}}
\multiput(255.34,733.00)(12.000,17.679){2}{\rule{0.800pt}{0.800pt}}
\multiput(270.41,754.00)(0.511,0.852){17}{\rule{0.123pt}{1.533pt}}
\multiput(267.34,754.00)(12.000,16.817){2}{\rule{0.800pt}{0.767pt}}
\multiput(282.41,774.00)(0.509,0.781){19}{\rule{0.123pt}{1.431pt}}
\multiput(279.34,774.00)(13.000,17.030){2}{\rule{0.800pt}{0.715pt}}
\multiput(295.41,794.00)(0.511,0.807){17}{\rule{0.123pt}{1.467pt}}
\multiput(292.34,794.00)(12.000,15.956){2}{\rule{0.800pt}{0.733pt}}
\multiput(307.41,813.00)(0.511,0.807){17}{\rule{0.123pt}{1.467pt}}
\multiput(304.34,813.00)(12.000,15.956){2}{\rule{0.800pt}{0.733pt}}
\multiput(319.41,832.00)(0.509,0.740){19}{\rule{0.123pt}{1.369pt}}
\multiput(316.34,832.00)(13.000,16.158){2}{\rule{0.800pt}{0.685pt}}
\multiput(332.41,851.00)(0.511,0.762){17}{\rule{0.123pt}{1.400pt}}
\multiput(329.34,851.00)(12.000,15.094){2}{\rule{0.800pt}{0.700pt}}
\multiput(344.41,869.00)(0.511,0.762){17}{\rule{0.123pt}{1.400pt}}
\multiput(341.34,869.00)(12.000,15.094){2}{\rule{0.800pt}{0.700pt}}
\multiput(356.41,887.00)(0.511,0.717){17}{\rule{0.123pt}{1.333pt}}
\multiput(353.34,887.00)(12.000,14.233){2}{\rule{0.800pt}{0.667pt}}
\multiput(368.41,904.00)(0.509,0.657){19}{\rule{0.123pt}{1.246pt}}
\multiput(365.34,904.00)(13.000,14.414){2}{\rule{0.800pt}{0.623pt}}
\multiput(381.41,921.00)(0.511,0.671){17}{\rule{0.123pt}{1.267pt}}
\multiput(378.34,921.00)(12.000,13.371){2}{\rule{0.800pt}{0.633pt}}
\multiput(393.41,937.00)(0.511,0.671){17}{\rule{0.123pt}{1.267pt}}
\multiput(390.34,937.00)(12.000,13.371){2}{\rule{0.800pt}{0.633pt}}
\multiput(405.41,953.00)(0.509,0.574){19}{\rule{0.123pt}{1.123pt}}
\multiput(402.34,953.00)(13.000,12.669){2}{\rule{0.800pt}{0.562pt}}
\multiput(418.41,968.00)(0.511,0.626){17}{\rule{0.123pt}{1.200pt}}
\multiput(415.34,968.00)(12.000,12.509){2}{\rule{0.800pt}{0.600pt}}
\multiput(430.41,983.00)(0.511,0.626){17}{\rule{0.123pt}{1.200pt}}
\multiput(427.34,983.00)(12.000,12.509){2}{\rule{0.800pt}{0.600pt}}
\multiput(442.41,998.00)(0.511,0.581){17}{\rule{0.123pt}{1.133pt}}
\multiput(439.34,998.00)(12.000,11.648){2}{\rule{0.800pt}{0.567pt}}
\multiput(453.00,1013.41)(0.492,0.509){19}{\rule{1.000pt}{0.123pt}}
\multiput(453.00,1010.34)(10.924,13.000){2}{\rule{0.500pt}{0.800pt}}
\multiput(467.41,1025.00)(0.511,0.536){17}{\rule{0.123pt}{1.067pt}}
\multiput(464.34,1025.00)(12.000,10.786){2}{\rule{0.800pt}{0.533pt}}
\multiput(479.41,1038.00)(0.511,0.536){17}{\rule{0.123pt}{1.067pt}}
\multiput(476.34,1038.00)(12.000,10.786){2}{\rule{0.800pt}{0.533pt}}
\multiput(490.00,1052.41)(0.536,0.511){17}{\rule{1.067pt}{0.123pt}}
\multiput(490.00,1049.34)(10.786,12.000){2}{\rule{0.533pt}{0.800pt}}
\multiput(503.00,1064.41)(0.491,0.511){17}{\rule{1.000pt}{0.123pt}}
\multiput(503.00,1061.34)(9.924,12.000){2}{\rule{0.500pt}{0.800pt}}
\multiput(515.00,1076.40)(0.539,0.512){15}{\rule{1.073pt}{0.123pt}}
\multiput(515.00,1073.34)(9.774,11.000){2}{\rule{0.536pt}{0.800pt}}
\multiput(527.00,1087.40)(0.539,0.512){15}{\rule{1.073pt}{0.123pt}}
\multiput(527.00,1084.34)(9.774,11.000){2}{\rule{0.536pt}{0.800pt}}
\multiput(539.00,1098.40)(0.589,0.512){15}{\rule{1.145pt}{0.123pt}}
\multiput(539.00,1095.34)(10.623,11.000){2}{\rule{0.573pt}{0.800pt}}
\multiput(552.00,1109.40)(0.599,0.514){13}{\rule{1.160pt}{0.124pt}}
\multiput(552.00,1106.34)(9.592,10.000){2}{\rule{0.580pt}{0.800pt}}
\multiput(564.00,1119.40)(0.674,0.516){11}{\rule{1.267pt}{0.124pt}}
\multiput(564.00,1116.34)(9.371,9.000){2}{\rule{0.633pt}{0.800pt}}
\multiput(576.00,1128.40)(0.674,0.516){11}{\rule{1.267pt}{0.124pt}}
\multiput(576.00,1125.34)(9.371,9.000){2}{\rule{0.633pt}{0.800pt}}
\multiput(588.00,1137.40)(0.737,0.516){11}{\rule{1.356pt}{0.124pt}}
\multiput(588.00,1134.34)(10.186,9.000){2}{\rule{0.678pt}{0.800pt}}
\multiput(601.00,1146.40)(0.774,0.520){9}{\rule{1.400pt}{0.125pt}}
\multiput(601.00,1143.34)(9.094,8.000){2}{\rule{0.700pt}{0.800pt}}
\multiput(613.00,1154.40)(0.774,0.520){9}{\rule{1.400pt}{0.125pt}}
\multiput(613.00,1151.34)(9.094,8.000){2}{\rule{0.700pt}{0.800pt}}
\multiput(625.00,1162.40)(1.000,0.526){7}{\rule{1.686pt}{0.127pt}}
\multiput(625.00,1159.34)(9.501,7.000){2}{\rule{0.843pt}{0.800pt}}
\multiput(638.00,1169.40)(0.913,0.526){7}{\rule{1.571pt}{0.127pt}}
\multiput(638.00,1166.34)(8.738,7.000){2}{\rule{0.786pt}{0.800pt}}
\multiput(650.00,1176.39)(1.132,0.536){5}{\rule{1.800pt}{0.129pt}}
\multiput(650.00,1173.34)(8.264,6.000){2}{\rule{0.900pt}{0.800pt}}
\multiput(662.00,1182.39)(1.132,0.536){5}{\rule{1.800pt}{0.129pt}}
\multiput(662.00,1179.34)(8.264,6.000){2}{\rule{0.900pt}{0.800pt}}
\multiput(674.00,1188.38)(1.768,0.560){3}{\rule{2.280pt}{0.135pt}}
\multiput(674.00,1185.34)(8.268,5.000){2}{\rule{1.140pt}{0.800pt}}
\multiput(687.00,1193.38)(1.600,0.560){3}{\rule{2.120pt}{0.135pt}}
\multiput(687.00,1190.34)(7.600,5.000){2}{\rule{1.060pt}{0.800pt}}
\multiput(699.00,1198.38)(1.600,0.560){3}{\rule{2.120pt}{0.135pt}}
\multiput(699.00,1195.34)(7.600,5.000){2}{\rule{1.060pt}{0.800pt}}
\put(711,1202.34){\rule{2.800pt}{0.800pt}}
\multiput(711.00,1200.34)(7.188,4.000){2}{\rule{1.400pt}{0.800pt}}
\put(724,1206.34){\rule{2.600pt}{0.800pt}}
\multiput(724.00,1204.34)(6.604,4.000){2}{\rule{1.300pt}{0.800pt}}
\put(736,1209.84){\rule{2.891pt}{0.800pt}}
\multiput(736.00,1208.34)(6.000,3.000){2}{\rule{1.445pt}{0.800pt}}
\put(748,1212.34){\rule{2.891pt}{0.800pt}}
\multiput(748.00,1211.34)(6.000,2.000){2}{\rule{1.445pt}{0.800pt}}
\put(760,1214.84){\rule{3.132pt}{0.800pt}}
\multiput(760.00,1213.34)(6.500,3.000){2}{\rule{1.566pt}{0.800pt}}
\put(773,1217.34){\rule{2.891pt}{0.800pt}}
\multiput(773.00,1216.34)(6.000,2.000){2}{\rule{1.445pt}{0.800pt}}
\put(785,1218.84){\rule{2.891pt}{0.800pt}}
\multiput(785.00,1218.34)(6.000,1.000){2}{\rule{1.445pt}{0.800pt}}
\put(797,1219.84){\rule{3.132pt}{0.800pt}}
\multiput(797.00,1219.34)(6.500,1.000){2}{\rule{1.566pt}{0.800pt}}
\put(846,1219.84){\rule{3.132pt}{0.800pt}}
\multiput(846.00,1220.34)(6.500,-1.000){2}{\rule{1.566pt}{0.800pt}}
\put(859,1218.84){\rule{2.891pt}{0.800pt}}
\multiput(859.00,1219.34)(6.000,-1.000){2}{\rule{1.445pt}{0.800pt}}
\put(871,1217.34){\rule{2.891pt}{0.800pt}}
\multiput(871.00,1218.34)(6.000,-2.000){2}{\rule{1.445pt}{0.800pt}}
\put(883,1214.84){\rule{3.132pt}{0.800pt}}
\multiput(883.00,1216.34)(6.500,-3.000){2}{\rule{1.566pt}{0.800pt}}
\put(896,1212.34){\rule{2.891pt}{0.800pt}}
\multiput(896.00,1213.34)(6.000,-2.000){2}{\rule{1.445pt}{0.800pt}}
\put(908,1209.84){\rule{2.891pt}{0.800pt}}
\multiput(908.00,1211.34)(6.000,-3.000){2}{\rule{1.445pt}{0.800pt}}
\put(920,1206.34){\rule{2.600pt}{0.800pt}}
\multiput(920.00,1208.34)(6.604,-4.000){2}{\rule{1.300pt}{0.800pt}}
\put(932,1202.34){\rule{2.800pt}{0.800pt}}
\multiput(932.00,1204.34)(7.188,-4.000){2}{\rule{1.400pt}{0.800pt}}
\multiput(945.00,1200.06)(1.600,-0.560){3}{\rule{2.120pt}{0.135pt}}
\multiput(945.00,1200.34)(7.600,-5.000){2}{\rule{1.060pt}{0.800pt}}
\multiput(957.00,1195.06)(1.600,-0.560){3}{\rule{2.120pt}{0.135pt}}
\multiput(957.00,1195.34)(7.600,-5.000){2}{\rule{1.060pt}{0.800pt}}
\multiput(969.00,1190.06)(1.768,-0.560){3}{\rule{2.280pt}{0.135pt}}
\multiput(969.00,1190.34)(8.268,-5.000){2}{\rule{1.140pt}{0.800pt}}
\multiput(982.00,1185.07)(1.132,-0.536){5}{\rule{1.800pt}{0.129pt}}
\multiput(982.00,1185.34)(8.264,-6.000){2}{\rule{0.900pt}{0.800pt}}
\multiput(994.00,1179.07)(1.132,-0.536){5}{\rule{1.800pt}{0.129pt}}
\multiput(994.00,1179.34)(8.264,-6.000){2}{\rule{0.900pt}{0.800pt}}
\multiput(1006.00,1173.08)(0.913,-0.526){7}{\rule{1.571pt}{0.127pt}}
\multiput(1006.00,1173.34)(8.738,-7.000){2}{\rule{0.786pt}{0.800pt}}
\multiput(1018.00,1166.08)(1.000,-0.526){7}{\rule{1.686pt}{0.127pt}}
\multiput(1018.00,1166.34)(9.501,-7.000){2}{\rule{0.843pt}{0.800pt}}
\multiput(1031.00,1159.08)(0.774,-0.520){9}{\rule{1.400pt}{0.125pt}}
\multiput(1031.00,1159.34)(9.094,-8.000){2}{\rule{0.700pt}{0.800pt}}
\multiput(1043.00,1151.08)(0.774,-0.520){9}{\rule{1.400pt}{0.125pt}}
\multiput(1043.00,1151.34)(9.094,-8.000){2}{\rule{0.700pt}{0.800pt}}
\multiput(1055.00,1143.08)(0.737,-0.516){11}{\rule{1.356pt}{0.124pt}}
\multiput(1055.00,1143.34)(10.186,-9.000){2}{\rule{0.678pt}{0.800pt}}
\multiput(1068.00,1134.08)(0.674,-0.516){11}{\rule{1.267pt}{0.124pt}}
\multiput(1068.00,1134.34)(9.371,-9.000){2}{\rule{0.633pt}{0.800pt}}
\multiput(1080.00,1125.08)(0.674,-0.516){11}{\rule{1.267pt}{0.124pt}}
\multiput(1080.00,1125.34)(9.371,-9.000){2}{\rule{0.633pt}{0.800pt}}
\multiput(1092.00,1116.08)(0.599,-0.514){13}{\rule{1.160pt}{0.124pt}}
\multiput(1092.00,1116.34)(9.592,-10.000){2}{\rule{0.580pt}{0.800pt}}
\multiput(1104.00,1106.08)(0.589,-0.512){15}{\rule{1.145pt}{0.123pt}}
\multiput(1104.00,1106.34)(10.623,-11.000){2}{\rule{0.573pt}{0.800pt}}
\multiput(1117.00,1095.08)(0.539,-0.512){15}{\rule{1.073pt}{0.123pt}}
\multiput(1117.00,1095.34)(9.774,-11.000){2}{\rule{0.536pt}{0.800pt}}
\multiput(1129.00,1084.08)(0.539,-0.512){15}{\rule{1.073pt}{0.123pt}}
\multiput(1129.00,1084.34)(9.774,-11.000){2}{\rule{0.536pt}{0.800pt}}
\multiput(1141.00,1073.08)(0.491,-0.511){17}{\rule{1.000pt}{0.123pt}}
\multiput(1141.00,1073.34)(9.924,-12.000){2}{\rule{0.500pt}{0.800pt}}
\multiput(1153.00,1061.08)(0.536,-0.511){17}{\rule{1.067pt}{0.123pt}}
\multiput(1153.00,1061.34)(10.786,-12.000){2}{\rule{0.533pt}{0.800pt}}
\multiput(1167.41,1046.57)(0.511,-0.536){17}{\rule{0.123pt}{1.067pt}}
\multiput(1164.34,1048.79)(12.000,-10.786){2}{\rule{0.800pt}{0.533pt}}
\multiput(1179.41,1033.57)(0.511,-0.536){17}{\rule{0.123pt}{1.067pt}}
\multiput(1176.34,1035.79)(12.000,-10.786){2}{\rule{0.800pt}{0.533pt}}
\multiput(1190.00,1023.08)(0.492,-0.509){19}{\rule{1.000pt}{0.123pt}}
\multiput(1190.00,1023.34)(10.924,-13.000){2}{\rule{0.500pt}{0.800pt}}
\multiput(1204.41,1007.30)(0.511,-0.581){17}{\rule{0.123pt}{1.133pt}}
\multiput(1201.34,1009.65)(12.000,-11.648){2}{\rule{0.800pt}{0.567pt}}
\multiput(1216.41,993.02)(0.511,-0.626){17}{\rule{0.123pt}{1.200pt}}
\multiput(1213.34,995.51)(12.000,-12.509){2}{\rule{0.800pt}{0.600pt}}
\multiput(1228.41,978.02)(0.511,-0.626){17}{\rule{0.123pt}{1.200pt}}
\multiput(1225.34,980.51)(12.000,-12.509){2}{\rule{0.800pt}{0.600pt}}
\multiput(1240.41,963.34)(0.509,-0.574){19}{\rule{0.123pt}{1.123pt}}
\multiput(1237.34,965.67)(13.000,-12.669){2}{\rule{0.800pt}{0.562pt}}
\multiput(1253.41,947.74)(0.511,-0.671){17}{\rule{0.123pt}{1.267pt}}
\multiput(1250.34,950.37)(12.000,-13.371){2}{\rule{0.800pt}{0.633pt}}
\multiput(1265.41,931.74)(0.511,-0.671){17}{\rule{0.123pt}{1.267pt}}
\multiput(1262.34,934.37)(12.000,-13.371){2}{\rule{0.800pt}{0.633pt}}
\multiput(1277.41,915.83)(0.509,-0.657){19}{\rule{0.123pt}{1.246pt}}
\multiput(1274.34,918.41)(13.000,-14.414){2}{\rule{0.800pt}{0.623pt}}
\multiput(1290.41,898.47)(0.511,-0.717){17}{\rule{0.123pt}{1.333pt}}
\multiput(1287.34,901.23)(12.000,-14.233){2}{\rule{0.800pt}{0.667pt}}
\multiput(1302.41,881.19)(0.511,-0.762){17}{\rule{0.123pt}{1.400pt}}
\multiput(1299.34,884.09)(12.000,-15.094){2}{\rule{0.800pt}{0.700pt}}
\multiput(1314.41,863.19)(0.511,-0.762){17}{\rule{0.123pt}{1.400pt}}
\multiput(1311.34,866.09)(12.000,-15.094){2}{\rule{0.800pt}{0.700pt}}
\multiput(1326.41,845.32)(0.509,-0.740){19}{\rule{0.123pt}{1.369pt}}
\multiput(1323.34,848.16)(13.000,-16.158){2}{\rule{0.800pt}{0.685pt}}
\multiput(1339.41,825.91)(0.511,-0.807){17}{\rule{0.123pt}{1.467pt}}
\multiput(1336.34,828.96)(12.000,-15.956){2}{\rule{0.800pt}{0.733pt}}
\multiput(1351.41,806.91)(0.511,-0.807){17}{\rule{0.123pt}{1.467pt}}
\multiput(1348.34,809.96)(12.000,-15.956){2}{\rule{0.800pt}{0.733pt}}
\multiput(1363.41,788.06)(0.509,-0.781){19}{\rule{0.123pt}{1.431pt}}
\multiput(1360.34,791.03)(13.000,-17.030){2}{\rule{0.800pt}{0.715pt}}
\multiput(1376.41,767.63)(0.511,-0.852){17}{\rule{0.123pt}{1.533pt}}
\multiput(1373.34,770.82)(12.000,-16.817){2}{\rule{0.800pt}{0.767pt}}
\multiput(1388.41,747.36)(0.511,-0.897){17}{\rule{0.123pt}{1.600pt}}
\multiput(1385.34,750.68)(12.000,-17.679){2}{\rule{0.800pt}{0.800pt}}
\multiput(1400.41,726.36)(0.511,-0.897){17}{\rule{0.123pt}{1.600pt}}
\multiput(1397.34,729.68)(12.000,-17.679){2}{\rule{0.800pt}{0.800pt}}
\multiput(1412.41,705.55)(0.509,-0.864){19}{\rule{0.123pt}{1.554pt}}
\multiput(1409.34,708.77)(13.000,-18.775){2}{\rule{0.800pt}{0.777pt}}
\multiput(1425.41,683.08)(0.511,-0.943){17}{\rule{0.123pt}{1.667pt}}
\multiput(1422.34,686.54)(12.000,-18.541){2}{\rule{0.800pt}{0.833pt}}
\put(810.0,1222.0){\rule[-0.400pt]{8.672pt}{0.800pt}}
\end{picture}
\caption{Mean field ground-state energy $(3.12)$ of the non-Abelian
anyon fluid, plotted as a function
of $\widetilde\rho_0$, for one value of $\kt$ in each phase.}
\end{figure}


The general feature of the non-Abelian problem is the presence of anyonic
particles with both isospin charges and opposite contributions to the
average iso-magnetic field $\widetilde{B}_0$.
If the two couplings have the same sign, the minimal energy ground-state
configuration has equal populations of isospin up and down particles
$ ( n_{+} = n_{-} ) $. This is the phase (i). In the phase (ii),
the difference of populations depends on the ratio $\kt/ \k$. In these
two cases, the non-Abelian mean field approximation is not self-consistent:
in the first one, $\widetilde{\rho}_0 = 0$ does not reproduce (\ref{nmf});
in the second case, the non-vanishing value of $\widetilde{\rho}_0$ is
achieved in the singular limit $n_{+}\to \infty$ and vanishing magnetic
field in the corresponding Landau level $ ( B + B_3 ) \to 0 $.
This case is not further analyzed here.
These two phases can be continuously connected to the theory with
$SU(2)$ gauge interaction only, by letting $1/|k| \ll 1/|\kt |$.
In particular, we find that the non-Abelian mean-field approximation
(\ref{nmf}) is not consistent for Chern-Simons theories with symmetry $SU(2)$
only, or for other semi-simple Lie algebras.
This $SU(2)$ invariant anyon fluid ground state is actually P and T
invariant, due to the vanishing of the iso-magnetic field (although the
fluctuations are not invariant). Similar models with a pair of oppositely
charged anyons have been introduced \cite{disass}, for explaining the lack
of P and T violations in the high-temperature superconductivity.
These models cannot be analyzed within this mean field approximation.

Here we shall discuss the phase (iii), where the $U(1)$ interaction is
dominating the $SU(2)$ one. Actually, this phase is continuously
connected with the previous Abelian model by letting $1/|\kt | \ll 1/|\k | $.
In this phase,
the lowest Landau level with field $| B_0 - \widetilde{B}_0 / 2 |$ is
empty, $ n_{-} = 0$, and has higher energy than the other Landau level
with field $| B_0 + \tilde{B}_0 / 2 |$, which is populated. A gap makes
stable this mean-field ground state against $( n_{+}, n_{-} )$
fluctuations.
Note that the non-Abelian interaction lowers the ground-state energy of
the pure Abelian theory (\ref{gse}), due to the cancellation mechanism
discussed above.

\subsection{ U(2) current algebra}

The analysis of low-energy, quadratic fluctuations around the non-Abelian
mean-field ground state is similar to the Abelian case in section
(2.3): we must derive the non-Abelian current algebra and rewrite the
Hamiltonian (\ref{nham}) in terms of currents.
The non-Abelian Chern-Simon field can be solved completely in terms
of the matter density at equal time by using the radial gauge \cite{rad},
\beq
x^i {\cal A}_i ( x ) = 0 \ . \label{rgu}
\eeq
These gauge conditions eliminate the commutator of non-Abelian gauge
fields appearing in the Gauss law (\ref{ngau}),
which can be solved as in the Abelian case (\ref{inv}),
\begin{eqnarray}
A_0^a & = & - \frac{1}{\kt }\ \frac{1}{x \cdot \partial}\
\epsilon_{ij} x^i J_j^a \ ,\nonumber \\
A_i^a & = & \frac{1}{\kt }\ \frac{1}{1+ x \cdot \partial}\
\epsilon_{ij} x^j \rho^a \ ,
\label{ninv} \end{eqnarray}
where the operators $(x\cdot\partial)^{-1}$ will be better defined afterwards.
Actually, in $(2+1)$-dimensions, the solution of the Gauss law is possible in
any axial gauge $ n^\mu A_\mu = 0 $ , at the expenses of breaking
either rotation or translation invariance.
In our problem, it is preferable to maintain explicit rotation invariance,
because the Chern-Simons interaction is chiral.

The non-Abelian current algebra can be obtained again by quantizing the
bosonic matter field $\Psi_i (x)$ only.
The commutation relations between the
gauge-invariant currents $( \rho, J_i )$ are still independent of the
Chern-Simons coupling and are given by the eqs. (\ref{cr}). The
commutation relations of $\rho^a$ are ,
\begin{eqnarray}
 \left[ { \hat{\rho} }^a ( {\bf x} ) , \hat{\rho}  ( {\bf y} ) \right]
& = & 0 \ ,\nonumber \\
 \left[ { \hat{\rho} }^a ( {\bf x} ) , { \hat{J} }_i ( {\bf y} ) \right]
& = & \frac{1}{im}\ {\partial\over\partial x^i}
\left(\delta({\bf x} - {\bf y})\rho^a({\bf x})\right)
+{1\over im\kt}\ \epsilon_{ij} \ y^j \epsilon_{abc} \rho^b ({\bf y})\
 \left( \frac{1}{ 2 + x \cdot \partial } \rho^c ({\bf x})
\right) \nonumber\\
&& - {1\over im\kt} \epsilon_{ij} \ x^j \epsilon_{abc} \rho^b ({\bf x})
\delta ({\bf x}-{\bf y}) \
\left( \frac{1}{ 2 + y \cdot \partial } \rho^c ({\bf y})  \right)
\ ,\nonumber \\
 \left[ { \hat{\rho} }^a ( {\bf x} ) , {\hat{\rho}}^b ( {\bf y} ) \right]
& = & i \ \epsilon_{abc}
\ \rho^c ( {\bf x} )\ \delta ( {\bf x} - {\bf y} ) \ .
\label{ncr} \end{eqnarray}
We use some algebraic identities relating the matter
and isospin ( $1/2 $ ) currents, which follows from
the completeness of the basis of $(2 \times 2)$ isospin matrices.
These are obtained by ``gauging'' the identities found in ref.\cite{sharp}:
\beq
4 \rho^a \rho^a = \rho^2 \ , \qquad
\rho J^a_i = \rho^a J_i - \frac{1}{m} \epsilon_{abc} \rho^b
{( {\cal D}_i \rho )}^c
\label{messy}\eeq
Therefore, we can consider the currents $( \rho^a, J_i )$ as independent
variables and $( \rho, J^a_i )$ as dependent ones. In particular, we do not
need the explicit form of the commutators involving $ J^a_i $.

Next, we study the normal ordering of the Hamiltonian (\ref{nham}).
The analysis is similar to the Abelian case (eqs. (\ref{ord})-(\ref{sub}))
because the non-Abelian ground state has the same filling of the
lowest Landau level.  Therefore, we must add to (\ref{nham}) a term
proportional to
the $U(2)$ Bogomol'nyi identity, analogous to (\ref{bom}),
with coefficient $ \alpha = 1 $. We then find:
\beq
{\cal{H}} = \int \ d^2 {\bf x} \ {1\over 2m}\ \left[ {( D_i \Psi )}^{\dagger}
( D_i \Psi ) + i \epsilon_{ij} {(D_i \Psi)}^{\dagger} D_j \Psi -
\frac{1}{\kappa} \rho^2 - \frac{1}{\kt } {( \rho^a )}^2 \right] \ .
\eeq
This can be written in terms of the currents $( \rho^a , J_i )$,
by using the identities (\ref{messy}) and another one for
${(D_i \Psi)}^{\dagger} {(D_j \Psi )}$ \cite{sharp}:
\barr
{\cal H} = \int d^2 {\bf x}\ {1\over 2m}\ \left[
\frac{1}{\rho}\left({({\cal D}_i \rho )}^a {({\cal D}_i \rho)}^a + m^2 J_i J_i
\right) \right.
& - &  \epsilon^{ij}\frac{1}{\rho} \left( m ( \partial_i \rho ) J_j + 2
 \ \epsilon^{abc} \ \frac{ \rho^a }{ \rho } \ {({\cal D}_i \rho )}^b
 {({\cal D}_i \rho )}^c \right) \nonumber \\
& - & \left. \frac{1}{\kappa} \rho^2 - \frac{1}{\kt } {( \rho^a )}^2 \right] .
\label{nah}\earr

\subsection{Quadratic fluctuations}

We expand the current algebra (\ref{ncr}) and the Hamiltonian (\ref{nah})
to leading order by plugging in the expansions
\beq
\rho^a  \simeq \frac{\rho_0}{2} \delta^a_3 + {\hat{\rho}}^a ({\bf x}) \ ,
\qquad J ({\bf x}) \simeq \hat J ({\bf x}) \ , \nonumber
\eeq
\beq
A^a_i \simeq \frac{1}{\kt}\ \epsilon_{ij} x^j \left( \delta^a_3 \
\frac{\rho_0}{4} + \frac{1}{2 + x \cdot {\bf \partial}} \
{\hat{\rho}}^a \right) \ , \label{qff}
\eeq
where the operators with hat are much smaller that their mean
field values, ( see eqs. (\ref{flu})-(\ref{app2})).
The expansion of the Hamiltonian involves the covariant derivative of the
isospin density ,
\beq
{({\cal D}_i {\hat{\rho}})}^a \simeq \partial_i {\hat{\rho}}^a +
\frac{\rho_0}{2\kt } \ \epsilon_{ij} x^j \ \epsilon^{ab3} \Lambda \
\hat{\rho}^b \ , \qquad
 \Lambda = \frac{1}{ 2 + x \cdot\partial } - \frac{1}{2} \ .
\label{cde}\eeq
The quadratic Hamiltonian splits into two terms (we omit the hats from now on),
\beq
{\cal H}^{(2)} = {\cal H}^{(2)}_A \ [ \rho^3, J_i ] \ +\
{\cal H}^{(2)}_{NA} \ [ \rho^1 , \rho^2 ] \ . \label{nah2}
\eeq
The first term is
\beq
{\cal H}^{(2)}_A = \int \ d^2 {\bf x} \ \frac{1}{2m \rho_0}
\left[ {(\partial_i \rho^3 )}^2 + m^2 {( J_i )}^2 - 2 m \epsilon_{ij}
\partial_i \rho^3 J_j - \rho_0 \left( \frac{4}{\k} + \frac{1}{\kt } \right)
{(\rho^3)}^2 \right] \ ,
\label{hamA}\eeq
and is similar to the Abelian Hamiltonian (\ref{qua}). It describes local
density fluctuations of the isospin-up particles in the lowest Landau
level, without isospin flips.
The second term in the Hamiltonian is
\begin{eqnarray}
{\cal H}^{(2)}_{NA} = \int d^2 {\bf x} \ \frac{1}{2m \rho_0}
\left[ {(\partial_i \rho^\alpha )}^2 \right.& - &
\rho_0 \left( \frac{4}{\k} + \frac{1}{2\kt} \right) {(\rho^\alpha)}^2
- \frac{\rho_0}{\kt } \left( \epsilon_{ij} x^i \partial_j \rho^\alpha \right)
 \epsilon_{\alpha\beta}\ \Lambda\ \rho^\beta
\nonumber \\
& + & \left.
{ \left( \frac{\rho_0}{2 \kt } \right) }^2 x^2 ( \Lambda \rho^{\alpha} )
( \Lambda \rho^{\alpha} ) \right]  \ ,
\label{hamNA}\end{eqnarray}
where the indices $\alpha , \beta = 1, 2$ are a subset of the adjoint isospin
indices. This term describes the non-Abelian dynamics of isospin-flip
excitations.

The approximated $U(2)$ current algebra (\ref{ncr}) of the
independent variables $( { \rho }^a, J_i )$ also decouples
into two orthogonal subalgebras of $(\rho^3, J_i)$ and $(\rho^1, \rho^2)$,
respectively:
\barr
 \left[ { \rho }^3 ( {\bf x}) , { \rho }^3 ( {\bf y}) \right] & = &
 \left[ J_i ( {\bf x}) , J_j ( {\bf y}) \right] = 0 \ ,\nonumber\\
 \left[ { \rho }^3 ( {\bf x}) , J_i ( {\bf y}) \right] & = &
{\rho_0\over 2im}\ \partial_i \delta ({\bf x}-{\bf y}) \ , \label{acNA}
\earr
and
\barr
 \left[ { \rho }^{\alpha} ({\bf x}) , { \rho }^{\alpha} ({\bf y}) \right]
 & = &  0 \qquad \alpha = 1, 2  \ ,\nonumber\\
 \left[ {\rho }^1 ({\bf x}) , { \rho }^2 ({\bf y}) \right] & = &
i \frac{\rho_0}{2} \delta ( {\bf x} - {\bf y} ) \ . \label{acNA2}
\earr
Therefore, the two terms of the quadratic Hamiltonian (\ref{nah2}) are
decoupled at the quantum level.
The $(\rho^3, J_i)$ subalgebra is isomorphic to the $U(1)$
current algebra (\ref{app},\ref{app2}),
thus the analysis of section 2.3 can be completely repeated
for the Hamiltonian (\ref{hamA}). After the Bogoliubov
transformation, one obtains a massless mode with sound velocity depending on
both Chern-Simons coupling constants,
\beq
v_s = \frac{1}{m} \sqrt{\rho_0 \left( \frac{1}{|\k|} - \frac{1}{4\kt} \right)}
\ , \qquad \qquad \left( 0 < \frac{1}{\kt} < \frac{4}{|\k|} \ ,
\ \ \k <0 \right)\ .
\eeq
Therefore, the density fluctuations which do not change isospin
behaves as the Abelian ones, with sound velocity related to the
ground-state energy (\ref{nagse}). Their physical interpretation
will be discussed in section five.

In the next section, we shall find the spectrum of isospin-flip
fluctuations.
The $(\rho^1, \rho^2)$ approximate current algebra can be represented by
a non-relativistic bosonic field $\chi$ as follows,
\beq
\rho^1 = \frac{\sqrt{\rho_0}}{2} \left( \chi + \chi^{\dagger} \right),
\qquad \rho^2 = \frac{\sqrt{\rho_0}}{2} i \left( \chi^{\dagger} - \chi \right)
\ ,\qquad \left[ \chi ({\bf x}), \chi^{\dagger} ({\bf y}) \right] =
\delta ( {\bf x} -{\bf y} ) \ .
\label{cqu}\eeq
Their Hamiltonian can be expressed in terms of the integro-differential
operator $\Delta$ acting on $\chi$,
\begin{eqnarray}
{\cal H}^{(2)}_{NA} & = & \int d^2 {\bf x} \ \chi^{\dagger}\
{1\over 2m} \left[ - \partial^2 + c + \frac{\rho_0}{2\kt } ( J \Lambda +
\Lambda^{\dagger} J ) + {\left( \frac{\rho_0}{2\kt } \right)}^2
\Lambda^{\dagger} x^2 \Lambda \right] \chi \nonumber\\
& = & \int d^2 {\bf x} \ \chi^{\dagger}\ \Delta\ \chi
\label{hamch}\end{eqnarray}
where $J= - i \epsilon_{ij} x_j \partial_i $ is the angular momentum and
$ c = - \rho_0 \left( 4 /\kappa +  1 / 2\kt  \right) > 0 $,
is a positive constant. Note that this Hamiltonian is normal-ordered, thus
there is no need of the Bogoliubov transformation in this case; the spectrum
of these excitations is given by the eigenvalues of the Hermitean
operator $\Delta$.

\section{The spectrum of non-Abelian fluctuations }

\subsection{General properties and gauge invariance}

The radial gauge condition (\ref{rgu}) breaks translation invariance,
because it selects a preferred origin of the coordinates of the plane.
This also happens in the Landau levels, due to the choice of the
background potential $ A_i = \frac{B}{2} \epsilon_{ij} x^j $.
Clearly, both systems have translation invariance, which can be achieved
by combining translations with compensating gauge transformations,
the so-called {\it magnetic} translations.
Non-observable non-gauge invariant quantities, like the eigen-functions,
transform covariantly under magnetic translations, while physical quantities,
like the energy levels, are invariant.

The magnetic translation operators for the non-Abelian
Hamiltonian (\ref{hamch}) are obtained by exploiting its close analogy
with the Landau level Hamiltonian \cite{ctz}.
We introduce the eigen-functions $\psi_{E,\ell}$ for one-particle states
with energy $E$ and angular momentum $\ell$,
\beq
\chi({\bf x}) = \sum_{E, \ell}  \psi_{E, \ell} ({\bf x})\
c_{E, \ell} \ ,
\label{expan}\eeq
and we find that they satisfy the Schr\"odinger-like equation,
\barr
\Delta\ \psi_{E,l} & = & E \psi_{E,l} = \left( - \frac{1}{2m} \sum_{i=1}^2
D_i^{\dagger} D_i + \frac{c}{2} \right) \psi_{E,l} \nonumber \\
D_i & = & \partial_i - i \hat{A}_i = \partial_i - i \epsilon_{ij} x^j
\frac{\rho_0}{2\kt}\left( \frac{1}{2 + x \cdot \partial} -
{1\over 2} \right)\ ,
\label{sch} \earr
which is similar to the Landau-level one, were it not for the additional
non-local operators in the covariant derivatives.
In the Landau level problem, the gauge transformation which restores the gauge
condition after translation is $(B = 2)$,
\beq
x^{'}_i = x_i + b_i , \ \ \ \ A^{'}_i - A_i = \partial_i \theta , \ \ \
\theta = \epsilon_{ij} x^i b^j , \ \ \ A^{'}_i = \epsilon_{ij} ( x^j + b^j )
\label{tra} \eeq
The eigen-functions are translated and rotated
\barr
\psi^{'}_L ({\bf x}^{'}) & = & e^{-i \theta} \psi_L ( {\bf x} + {\rm b} )
\simeq ( 1 + b^i \pi_i^L+ O ( b^2 )  ) \psi_L ( {\bf x} )  \nonumber \\
\pi_i^L & = & \frac{\partial}{\partial x_i} + i \epsilon_{ij} x^j
\label{shi} \earr
while the Hamiltonian $H^L = -  {( \partial/\partial x^i
- i A_i )}^2/ 2m $ is covariant. This implies
\beq [ H^L , \pi_i^L ] = 0 \ .\label{hli}\eeq

In the non-Abelian problem (\ref{sch}), the compensating gauge
transformation from
$x^i A_i^a = 0 $ to $ ( x^i + b^i ) A^{' a}_i = 0 $, with $A_i^a$ given by
(\ref{qff}), is similarly found to be
\barr
A_i^{'a} &=& A_i^a + \partial_i \theta^a - \epsilon_{abc} \theta^b A_i^c
\nonumber \\
\theta^a  & = &  \frac{1}{\kt } \epsilon_{ij} x^i b^j
\left( \delta^a_3 \frac{\rho_0}{4} + \frac{1}{(1 + x \cdot \partial)
(2 + x \cdot \partial )} \rho^a \right) \label{tra2}
\earr
to leading order in the fluctuating densities $\rho^a$. Their approximate gauge
transformations are found to be
\barr
\rho^{'3} & = & \rho^3  \nonumber \\
\chi^{'}  & = & e^{-i \hat{\theta}} \chi \simeq \left[ 1 -
i   \epsilon_{ij} x^i b^j
\frac{\rho_0}{2\kt } \left(\frac{1}{2} -\frac{1}{(1 + x\cdot \partial)
(2 + x\cdot \partial )} \right) \right] \chi \ ,
\label{shi2} \earr
thus $\chi$ acquires an operator-valued phase.
One can show that the derivatives $D_i$ in (\ref{sch}) are ``covariant''
in the operator sense: $ D^{'}_i e^{-i\hat{\theta}} = e^{-i\theta} D_i $,
where $\theta$ is given by (\ref{tra}).
Thus, in analogy with (\ref{shi},\ref{hli}), the operator
$\Delta$ commutes with the following magnetic translation operator,
\beq
\pi_i = \partial_i + i \epsilon_{ij} x^j \ \frac{\rho_0}{2 \kt }
\left( \frac{1}{2} - \frac{1}{(1+x\cdot\partial)(2+x\cdot\partial)} \right)
\ ,\qquad [ \Delta , \pi_i ] = 0 \ .
\label{hli2} \eeq
It is convenient to introduce the holomorphic components $\pi,
\pi^{\dagger}$,
\beq \pi = \frac{1}{2} ( \pi_1 - i \pi_2 ) =
\frac{\partial}{\partial z}  + \frac{\overline{z}}{2}
- z \frac{1}{(1+x\cdot\partial)(2+x\cdot\partial)} \label{hol} \eeq
(hereafter, the length scale given by the non-Abelian mean field
$\rho_0 /2\kt $ is put equal to 2 and the mass $m=1$).
The operators $\pi$ and $\pi^{\dagger}$ satisfy the same algebra as their
simpler Landau-level analogues, because the non-local interaction mediated by
the Chern-Simons field is translation invariant,
\beq [ \pi, \pi^{\dagger} ] = 1 , \ \ \ \ [ J, \pi ] = - \pi , \ \ \ \
\label{alg} \eeq
These relations imply that the spectrum of $\Delta$ is infinitely
degenerate in angular momentum, because
$\pi \psi_{E,\ell} \propto \psi_{E, \ell - 1} $ or $\pi \psi_{E, \ell} = 0 $.
Another analogous identity is
\beq 2 \pi^{\dagger} \pi = \left( \Delta - \frac{c}{2} - 1 \right) +
2 J \label{idd} \eeq
which shows that $\pi$ is invertible apart from a special line in the $(E,J)$
plane of the spectrum, which we shall discuss later on. Away from this line,
$\pi^\dagger$ can be used to generate the eigen-functions of arbitrary positive
$\ell$ starting from any given value, say from $\ell = 0$.

As in the Landau-level problem, the operator $\Delta$ can be put into a
manifest positive-definite form, by introducing the holomorphic components
of the covariant derivatives $D_i$ (\ref{sch}), as follows,
\begin{eqnarray}
\Delta & = & 2 a^{\dagger} a + \frac{c}{2} - 1 \ ,\nonumber \\
a & = & \frac{\partial}{\partial \bar{z}}
+ z \left( \frac{1}{2} - \frac{1}{2 + x\cdot\partial} \right), \ \ \
a^{\dagger} = - \frac{\partial}{\partial z}
+ \left( \frac{1}{2} + \frac{1}{x\cdot\partial} \right) \bar{z}
\label{sch2} \end{eqnarray}
Note, however, that the eigenvalue problem is much harder, because the
operators
$ a , a^{\dagger} $ do not satisfy the simple harmonic oscillator algebra.
The form (\ref{sch2}) for $\Delta$ allows to put a positive lower bound on the
energy,
\beq E \geq \frac{c}{2} - \frac{\rho_0}{4\kt}
\equiv \rho_0 \left( \frac{2}{|\k|} - \frac{1}
{2 \kt } \right) > 0 \label{epo} \eeq
which implies that the spectrum for the non-Abelian fluctuations is positive
definite and has a mass gap. In the following discussion, we shall show that
this bound is saturated and explain its physical origin.

\subsection{Physical interpretation of the spectrum}

In the previous section, we have shown that:

i) the operator $\Delta$ looks like a non-local deformation of the Landau level
problem of electrons in the average effective magnetic field $B_{\rm eff} =
\rho_0 / {2\kt }$;

ii) the spectrum has a gap $ M = \rho_0 ( 2 / |\k| - 1 / 2\kt  ) $,
proportional to both couplings.

Let us try to explain these results in simple terms before entering in the more
technical analysis of the eigenvalue problem.

The zero-th order mean field approximation has provided us with the physical
picture of two effective Landau level structures for up and down isospin
particles, with effective magnetic fields
$\rho_0 ( 1/|\k| - 1 / 4\kt )$ and $\rho_0 ( 1 / |\k| + 1 / 4\kt )$,
respectively. The ground state, corresponding to the homogeneous filling
of the lowest isospin-up level, has the energy
${\cal E}_0 = \rho_0( 1 /|\k|- 1/4\kt)/2$ per particle (\ref{nagse}),
which is given by the expectation value of the two-body repulsive term,
\beq
\int d^2{\bf x} \ \frac{1}{2} \left( B \rho + B^a \rho^a \right)
= {\cal E}_0 \rho_0 A + 4 {\cal E}_0 \int d^2 {\bf x}
\left( \hat\rho^a \right)^2 \ ,
\label{twb} \eeq
in the Hamiltonian (\ref{nah}).
This repulsive interaction affects both the density fluctuations which are
isospin diagonal $ ( \rho^3 \propto a^{\dagger}_{\uparrow} a_{\uparrow} )$ and
isospin rotating ( $ \rho^1+i\rho^2 \propto \psi \propto
a^{\dagger}_{\downarrow} a_{\uparrow} $ ). The former are Bogoliubov
transformed, and therefore are gapless, with sound
velocity proportional to $\sqrt{{\cal E}_0}$. The latter excitations are not
transformed and then acquire the gap $ M = 4 {\cal E}_0 $ from (\ref{twb}).

In more physical terms, the isospin diagonal phonons are local fluctuations
in the filling of the lowest Landau level, which do not feel a net
magnetic field. On the other hand, the isospin rotating fluctuations are made
of individual isospin flips, which move one electron from the filled up-level
to an empty down-level.
This can be thought of as leading to two effects: the {\it hole}
in the filled up-level propagates as a phonon in a magnetic field, the
{\it magneto-phonon}, which is gapful \cite{doria}. The jump of the
electron to any empty down-level gives a discrete spectrum with steps
proportional to the non-Abelian magnetic field $\rho_0 / 2 \kt $.
Therefore, we expect that the operator $\Delta$ has a discrete, Landau-like
spectrum above the magneto-phonon gap $M$.

\subsection{Eigenvectors and eigenvalues}

After the separation of variables,
\beq
\psi_{E,\ell} ( r, \theta ) = e^{i \ell \theta } \psi_{E, \ell} ( r )
 \eeq
the eigenvalue problem (\ref{sch}) can be rewritten, for the radial part,
\begin{eqnarray} 2 ( E - \Delta ) \psi_{E,\ell}(r)& = & \left[ \frac{1}{r^2}
( {( r \partial_r )}^2 - \ell^2 ) + 2 \ell + 2 \lambda - r^2 + \right.
\nonumber \\
 &\ & + \left.  4 \frac{1}{r\partial_r} \left( r^2 + \ell\right)
\frac{1}{2 + r\partial_r} \right] \psi_{E,l} (r)\ ,
\label{rad} \end{eqnarray}
where we parametrized $ E = \lambda + c / 2 = \lambda + 1 + M $.
The operator in the first line of this equation is the Landau
Hamiltonian, which would yield the spectrum  $\lambda = 2 k + 1 , k \geq 0 $;
the second line is the additional non-local term.
It is convenient to transform (\ref{rad})  into a fourth-order differential
equation for the  reduced wave-function $ \psi = ( 2 + r \partial_r ) \varphi
$.
By multiplying also on the left by $( r^3 \partial_r )$, we obtain
\begin{eqnarray}
&\ & \left[ {(r\partial_r)}^4 + {(r\partial_r)}
( - r^4 + 2 ( \lambda + \ell ) r^2
- {\ell}^2 - 4 ) ( r \partial_r ) + 4 \ell ( \ell + r^2 ) \right] \varphi = 0
\ ,\nonumber \\
&\ & \psi = ( 2 + r \partial_r ) \varphi \ .
\label{foe} \end{eqnarray}
Although the operator $(2 + r \partial_r)$ has a non-trivial kernel, we shall
find that it is invertible in the subspace of physically (normalizable)
wave-functions, for which the two eigenvalue problems (\ref{rad}) and
(\ref{foe}) are actually equivalent.

The analysis of the characteristic  equations for the solutions of
(\ref{foe}) around $r = 0$
and $r = \infty $, leads to the following asymptotic behaviors,
\beq
\begin{array}{c}
r \rightarrow 0 \\ \hrulefill \\
  \begin{array}{rlcrl}
  \varphi & \simeq r^{|\ell|}\ , & \ & \psi & \simeq r^{|\ell|} \\
   \      & \simeq r^{-|\ell|}\ ,& \ &  \   & \simeq r^{-|\ell|} \\
   \      & \simeq r^2 \ ,       & \ &  \   & \simeq r^2        \\
   \      & \simeq r^{-2} \ ,    & \ &  \   & \simeq r^0
  \end{array} \ \ \
\end{array}
\begin{array}{c}
r \rightarrow\infty \\ \hrulefill \\
  \begin{array}{rlcrl}
  \varphi &\simeq {\rm e}^{-r^2/2}\ , & \ & \psi &\simeq {\rm e}^{-r^2/2} \\
  \       &\simeq {\rm e}^{r^2/2}\ ,  & \ & \    &\simeq {\rm e}^{r^2/2} \\
  \       &\simeq  r^0 \ ,            & \ & \    &\simeq r^0           \\
  \       &\simeq  r^{-4} \ ,         & \ & \    &\simeq r^{-4}
  \end{array}
\end{array}
\label{fab}\eeq
The first two behaviors for both $r\to 0$ and $r \to\infty$ in this table
are found in the Landau problem, while the last two ones are new.
Note that these four behaviors can also be obtained from the
integro-differential form (\ref{rad}), by introducing two constants
for the homogeneous solutions of the integral operators.

Since the $r \to\infty$ asymptotics of free waves are not found in
(\ref{fab}),
we conclude that the physical solutions are square-integrable and that the
spectrum is discrete. The integrable behaviors are $\psi \simeq
( r^{|\ell|} , r^0 , r^2  ) $ for $ r \rightarrow 0$ and $ \psi \simeq  (
e^{-r^2/2}, r^{-4} ) $ for $r \rightarrow \infty$.

Next, we analyze the action of the magnetic translation operator
$\pi$ (\ref{hol}).
The multiple action of $\pi$ generates all the integral operators
\beq
\frac{1}{m + r\partial_r}\ \psi (r) = \frac{1}{r^m}\
\frac{1}{r\partial_r}\  r^m\ \psi (r), \qquad  m \in {\rm Z}\ ,
\label{iop} \eeq
whose action should be well-defined and unique.
We first need the precise definition of these integral operators \cite{rad}:
\beq
F = \frac{1}{r\partial_r} f(r) = \int^1_a \frac{d\lambda}{\lambda}
\ f ( \lambda r ) \ .
\label{iac} \eeq
In this equation, the constant $a$ parametrizes the homogeneous solution of
$r\partial_r F= F $, which corresponds to the residual gauge freedom
within the radial gauge (\ref{rgu});
we take a global complete gauge fixing by setting
$a = \infty$ in (\ref{iac}), where $\psi $ and $\varphi $ vanish.
This choice leads to a consistent solution of the eigenvalue problem\footnote{
Here, we do not find any analogous of the obstruction to fixing completely
the axial gauge described in ref. [23].}.
This gauge choice enforces the physical condition that matter fluctuations
vanishing at infinity should not produce a gauge field at infinity.
In this gauge, the integral operators (\ref{iop}) are well defined on
wave-functions with asymptotics $ \psi \simeq {\rm e}^{-r^2/2}$
$(r\to\infty)$, but can be singular on $\psi \simeq r^{-4} $;
therefore, we neglect the latter type of solutions.
We have shown that gauge invariance (the action of $\pi$) imposes further
conditions on the physical solutions, which also
ensure the invertibility of the relation between $\phi$ and $\psi$.
The explicit action of the integral operators (\ref{iac}) on the basis
of polynomials times the exponential asymptotics is,
\beq
\frac{1}{r\partial_r} \ r^\beta e^{-r^2/2} \equiv
\int_\infty^1 \frac{d\lambda}{\lambda}\ {(\lambda r)}^\beta \
{\rm e}^{- \lambda^2 r^2/2} = - { {\rm e}^{-r^2/2}\over 2}\ r^\beta\
\Psi \left( 1, 1 + \frac{\beta}{2} ; \frac{r^2}{2} \right) \ ,
\label{asy} \eeq
where $\Psi$ is the confluent hypergeometric function vanishing at
$r=\infty$ \cite{bat}. This is also the incomplete gamma function
$\Gamma ( \beta/2,r^2/2 )$, which is polynomial for $\beta = 2, 4 , 6, ..., $
and has logarithmic terms for $\beta = 0, - 2, - 4, ... $.
Its expansion for $r \rightarrow 0$ is,
\beq
\frac{1}{r\partial_r}\ r^\beta {\rm e}^{-r^2/2} \
{\buildrel r\to 0 \over \longrightarrow}\ \left\{
\begin{array}{ll}
{\rm e}^{-r^2/2} \ \frac{r^\beta}{\beta}
\left( 1 + \frac{r^2}{2+\beta} + O ( r^4 ) \right)
- \Gamma ( \frac{\beta}{2} ) 2^{\beta/2 - 1}\ , & \beta \neq 0,-2,-4,..., \\
\frac{1}{2}\ \log r^2 + O ( 1 )\ , & \beta = 0, -2,-4,...
\end{array} \right.
\label{expa} \eeq

In conclusion, we accept integrable eigen-functions with asymptotics
$\psi \simeq {\rm e}^{-r^2/2}$, for $r\to \infty$, and
$\psi \simeq (r^0, r^2, r^{|\ell|}) $ for $ r \rightarrow 0 $.
The counting of free parameters is as follows: there are four physical
independent solutions in total (three at $ r \simeq 0 $ and one at
$ r \simeq \infty\ $), plus the energy, minus the four matching
conditions $ {(d/dr)}^n \psi ( r_0 ) $, $n = 0 , 1, 2, 3$,  at
$ 0 < r_0 < \infty $, and the normalization condition. This counting is
consistent with a unique physical solution and a discrete spectrum.
Actually, we shall find that the system of conditions is
over-determined ( $- 1$ free parameters ), because the logarithmic solution
corresponding to the degenerate exponents $\alpha = 0, 2 $, at $ r = 0 $,
will not be present.

\bigskip

For $\ell=0$, the subset of physical solutions of the differential equation
(\ref{foe}) are also solutions of the following second order equation,
\beq
\left[ {( r\partial_r )}^2 - r^4 + 2 \lambda r^2 - 4 \right] \phi = 0\ ,
\qquad \psi = ( 2 + r \partial_r )\ \varphi =
\frac{2+r\partial_r}{r\partial_r} \phi \ ,
\label{soe} \eeq
for the function $\phi$. Indeed, the asymptotic behaviors of this equation,
$\phi \simeq r^{\pm 2}\ ( r \rightarrow 0 )$ and
$ \phi \simeq {\rm e}^{\pm r^2/2}\ ( r \rightarrow \infty ) $ correspond
to $\psi \simeq r^0, r^2,\ $ and $\psi \simeq {\rm e}^{\pm r^2/2}$,
respectively.
The general solutions of (\ref{soe}) are readily found in terms of
confluent hypergeometric functions $\Psi$ and $\Phi$:
the unique regular solution at infinity is given by,
\beq
\phi = r^2\ {\rm e}^{-r^2/2}\ \left\{
\begin{array}{ll}
\Phi ( 1 - k , 3 ; r^2 ),  & \ \ k = 1, 2, 3, ..., \\
\Psi ( 1 - k , 3 ; r^2 ), & \ \ k \neq 1, 2, 3, ...,
\end{array} \right.
\label{hyp} \eeq
as a function of the energy parameter $ E = 2 k + 2 + M $.
The $\Phi$ solutions corresponds to the discrete spectrum
\beq E = \frac{\rho_0}{2\kt m} ( k + 1 ) + M \ ,
\qquad M= {\rho_0\over m}\left({2\over |\k|}-{1\over 2\kt}\right)\ ,
\qquad
k = 1, 2, 3, ..., \ .
\label{spe} \eeq
(Physical units were restored in (\ref{spe})).
Their wave-functions are polynomial, because $\Phi ( 1 - k, 3 ; r^2 )$
truncates at the $k$-th term and behaves as $\Phi\simeq r^2\ (r\to 0 )$.
The corresponding $\psi$ functions are also polynomial,
as shown by using (\ref{expa}). Therefore, these solutions are acceptable.

The second type of solutions exists for the complementary continuous range of
energy, and behaves as $\Phi \simeq r^{-2}$ for $ r \rightarrow 0$.
As a consequence, the corresponding $\psi$ functions are logarithmic
for $r \to 0$,
\barr
\psi(r) & =& \frac{2 + r\partial_r}{r\partial_r}
\frac{1}{\Gamma(1-k)}\left(\frac{1}{r^2} + k+1+ O(r^2 ) \right) e^{-r^2/2}
\nonumber\\
& {\buildrel r \rightarrow 0 \over \longrightarrow }&
- \frac{1}{2 \Gamma ( 1 - k ) } \ \log r^2 \ ,\qquad
k \neq 1, 2, 3, ...
\label{logg} \earr
Although the logarithmic behavior is square-integrable, it produces a
$\delta ( r )$ term in the r.h.s. of the eigenvalue equation (\ref{rad}),
due to $( \partial_i^2 + ... )\psi \propto \delta ( x )$,
which cannot be accepted. Another reason for
rejecting these solutions is that they are mapped by $\pi^{\dagger}$ and $\pi$
into non-integrable solutions $\psi\simeq r^{-|\ell|}$ $(r\to 0)$
with $\ell=\pm 1$, as shown in the appendix.

In conclusion, the spectrum for $\ell = 0$ is discrete and given by
(\ref{spe}). Let us add some remarks:

\noindent
i) The number of free parameters for the reduced second-order problem is
equal to zero, because there are two physical independent solutions, plus
the energy, minus two matching conditions and the normalization.
This is correct for a discrete spectrum and show no sign of the $\ell \neq 0$
over-determination mentioned before.

\noindent
ii) Eigenfunctions of (\ref{foe}) with asymptotic
$\psi \simeq r^{-4}\ ( r \to\infty)$
correspond for $\ell = 0$ to the solutions of the inhomogeneous equation
$ [ {( r \partial_r )}^2 - r^4 + 2 \lambda r^2 - 4 ] \phi = 1$.
In the appendix, we show that they behave as $ \psi = O ( \log r^2 ) (
r \rightarrow 0 ) $ and should be rejected by the same arguments given
for eq. (\ref{logg}).
This is another reason for discarding this type of solutions,
independent of (\ref{iop}).

\noindent
iii) The physical solutions (\ref{spe}) are of the form
$\psi = ( 2 + r\partial_r )\varphi$, with $\varphi$ regular for $r = 0$,
thus they have vanishing zero mode
\beq
\left. \int d^2 {\bf x} \ \psi\right\vert_{\ell=0}\
= 2 \pi \int_0^{\infty} dr \partial_r ( r^2 \varphi ) = 0
\label{zmd} \eeq
as required for density fluctuations around the mean field.

The $\ell=0$ spectrum (\ref{spe}) extends for $\ell \neq 0$ into Landau
-like levels, whose eigen-functions can be obtained in
principle by applying the $\pi$ and $\pi^{\dagger}$ magnetic translation
operators (\ref{hol},\ref{alg}) to the $ \ell = 0 $ functions.
For $\ell \neq 0$, we do not have a general explicit method of solution
and we made a case by case analysis.
The properties of the $\ell \neq 0$ solutions are summarized in fig. 2 and
will be briefly discussed hereafter, leaving the details to the appendix.
The allowed values of $\ell$ are bounded from below by the line $k = - \ell $,
because for these values $\pi$ is not invertible, by eq. (\ref{idd}),
and annihilates the physical wave functions.


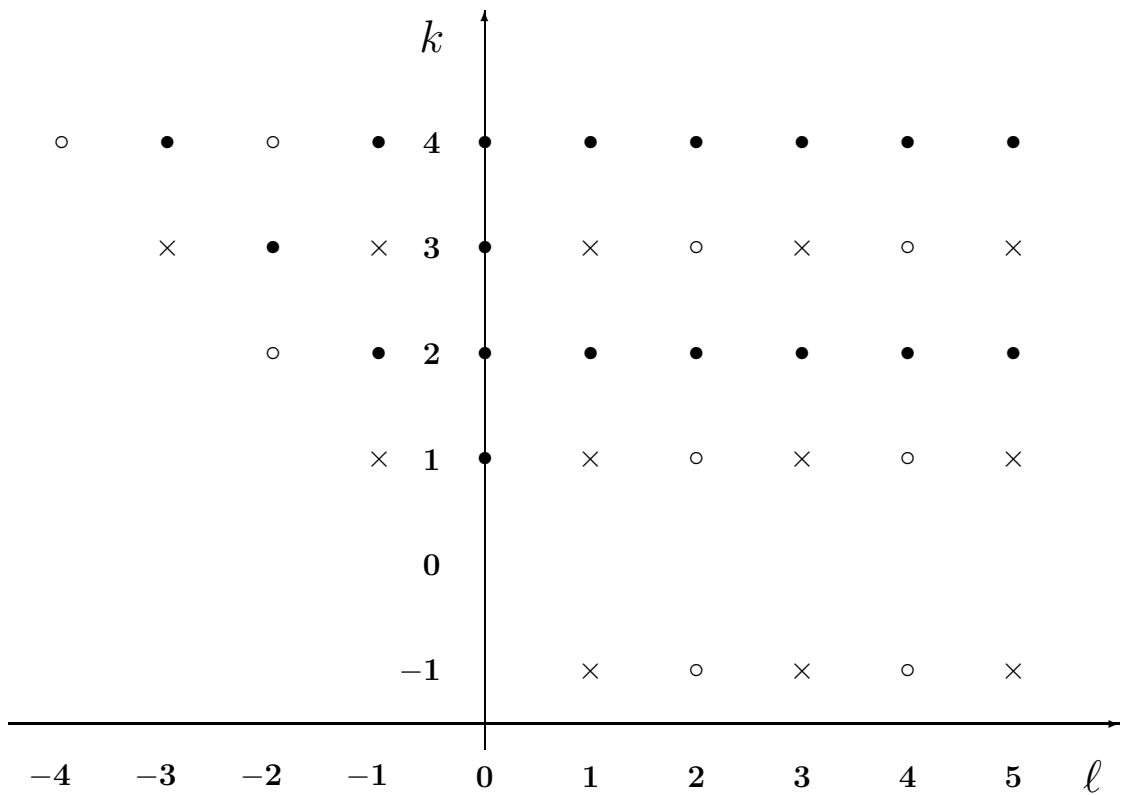
\begin{figure}
\unitlength=1.pt
\begin{picture}(440.00,400.00)(0.00,0.00)
\put(20.00,80.00){\vector(1,0){420.00}}
\put(200.00,70.00){\vector(0,1){280.00}}

\put(40.00,60.00){\makebox(0,0)[cc]{$\bf -4 \ \ $}}
\put(80.00,60.00){\makebox(0,0)[cc]{$\bf -3 \ \ $}}
\put(120.00,60.00){\makebox(0,0)[cc]{$\bf -2 \ \ $}}
\put(160.00,60.00){\makebox(0,0)[cc]{$\bf -1 \ \ $}}
\put(200.00,60.00){\makebox(0,0)[cc]{$\bf  0 $}}
\put(240.00,60.00){\makebox(0,0)[cc]{$\bf  1 $}}
\put(280.00,60.00){\makebox(0,0)[cc]{$\bf  2 $}}
\put(320.00,60.00){\makebox(0,0)[cc]{$\bf  3 $}}
\put(360.00,60.00){\makebox(0,0)[cc]{$\bf  4 $}}
\put(400.00,60.00){\makebox(0,0)[cc]{$\bf  5 $}}
\put(430.00,60.00){\makebox(0,0)[cc]{\Large $  \ell $}}

\put(180.00,100.00){\makebox(0,0)[cc]{$\bf -1 \ \ $}}
\put(180.00,140.00){\makebox(0,0)[cc]{$\bf  0 $}}
\put(180.00,180.00){\makebox(0,0)[cc]{$\bf  1 $}}
\put(180.00,220.00){\makebox(0,0)[cc]{$\bf  2 $}}
\put(180.00,260.00){\makebox(0,0)[cc]{$\bf  3 $}}
\put(180.00,300.00){\makebox(0,0)[cc]{$\bf  4 $}}
\put(180.00,340.00){\makebox(0,0)[cc]{\Large $ k $}}

\put(40.00,300.00){\makebox(0,0)[cc]{$ \circ $}}
\put(80.00,300.00){\makebox(0,0)[cc]{$ \bullet $}}
\put(120.00,300.00){\makebox(0,0)[cc]{$\circ $}}
\put(160.00,300.00){\makebox(0,0)[cc]{$\bullet $}}
\put(200.00,300.00){\makebox(0,0)[cc]{$\bullet $}}
\put(240.00,300.00){\makebox(0,0)[cc]{$\bullet $}}
\put(280.00,300.00){\makebox(0,0)[cc]{$\bullet $}}
\put(320.00,300.00){\makebox(0,0)[cc]{$\bullet $}}
\put(360.00,300.00){\makebox(0,0)[cc]{$\bullet $}}
\put(400.00,300.00){\makebox(0,0)[cc]{$\bullet $}}

\put(80.00,260.00){\makebox(0,0)[cc]{$ \times $}}
\put(120.00,260.00){\makebox(0,0)[cc]{$ \bullet $}}
\put(160.00,260.00){\makebox(0,0)[cc]{$ \times  $}}
\put(200.00,260.00){\makebox(0,0)[cc]{$ \bullet $}}
\put(240.00,260.00){\makebox(0,0)[cc]{$ \times  $}}
\put(280.00,260.00){\makebox(0,0)[cc]{$ \circ   $}}
\put(320.00,260.00){\makebox(0,0)[cc]{$ \times  $}}
\put(360.00,260.00){\makebox(0,0)[cc]{$ \circ   $}}
\put(400.00,260.00){\makebox(0,0)[cc]{$ \times  $}}

\put(120.00,220.00){\makebox(0,0)[cc]{$ \circ   $}}
\put(160.00,220.00){\makebox(0,0)[cc]{$ \bullet $}}
\put(200.00,220.00){\makebox(0,0)[cc]{$ \bullet $}}
\put(240.00,220.00){\makebox(0,0)[cc]{$ \bullet $}}
\put(280.00,220.00){\makebox(0,0)[cc]{$ \bullet $}}
\put(320.00,220.00){\makebox(0,0)[cc]{$ \bullet $}}
\put(360.00,220.00){\makebox(0,0)[cc]{$ \bullet $}}
\put(400.00,220.00){\makebox(0,0)[cc]{$ \bullet $}}

\put(160.00,180.00){\makebox(0,0)[cc]{$ \times  $}}
\put(200.00,180.00){\makebox(0,0)[cc]{$ \bullet $}}
\put(240.00,180.00){\makebox(0,0)[cc]{$ \times  $}}
\put(280.00,180.00){\makebox(0,0)[cc]{$ \circ   $}}
\put(320.00,180.00){\makebox(0,0)[cc]{$ \times  $}}
\put(360.00,180.00){\makebox(0,0)[cc]{$ \circ   $}}
\put(400.00,180.00){\makebox(0,0)[cc]{$ \times  $}}

\put(240.00,100.00){\makebox(0,0)[cc]{$ \times $}}
\put(280.00,100.00){\makebox(0,0)[cc]{$ \circ  $}}
\put(320.00,100.00){\makebox(0,0)[cc]{$ \times $}}
\put(360.00,100.00){\makebox(0,0)[cc]{$ \circ  $}}
\put(400.00,100.00){\makebox(0,0)[cc]{$ \times $}}

\end{picture}

\caption{Synopsis of the physical eigenstates in the angular momentum
$(\ell)$ versus energy $(k)$ plane (The meaning of symbols $(\bullet)$,
$(\circ)$ and $(\times)$ is given in the text).}
\end{figure}


Let us first note that the condition (\ref{zmd}) is trivially
satisfied by $\ell \neq 0$ eigen-functions, due to their angular dependence,
while it eliminates the zero mode for $\ell=0$.
This implies that another line of eigenstates should exist for $\ell > 0 $,
which is not connected to the previous $\ell = 0 $ states.
This is possible for the energy $ E = M, ( k = - 1 ) $, because $\pi$
is not invertible at $\ell = 1$, $\pi \psi_{-1,1} = 0 $, where the line stops.
Moreover, the bound (\ref{epo}) forbids states of lower energy. Indeed, such
eigenstates are found by explicit analysis of the differential equation
(\ref{foe}).

The general properties of $(\ell\neq 0)$ solutions are the following
(see fig. 2). There are polynomial solutions, which were found
by solving the four-term recursion relation with Mathematica \cite{wolf}.
These are of two types:

\noindent i) for $k\ge 2$ even, and $\ell \ge 0$,
\beq
\psi = e^{i \ell \theta} r^{|\ell|}
\left( a_0 + a_1 r^2 + ... + a_k r^{2k} \right)
e^{-r^2/2} \ ,\qquad ( \ {\rm points} \  ( \bullet ) ),
\label{pol}\eeq
ii) for $k\ge -1$ and $\ell\ge -k$, $\ell$ even,
\beq
\psi = e^{i \ell \theta} \left( b_0 + b_1 r^2 + ... + b_{k+\ell/2}
r^{2k + \ell} \right) e^{-r^2/2} \ ,\qquad ( \ {\rm points} \ ( \circ ) ),
\label{pol2} \eeq
Moreover, for odd $\ell$, there are non-polynomial solutions,
represented by crosses ($\times$) in fig. 2, which display a double
power expansion:
\beq
\psi = e^{i \ell \theta}\ \left\{ r^{|\ell |}
\left( a_0 + a_1 r^2 + \cdots \right)+
 b_0 + b_1 r^2 + \cdots \right\} e^{-r^2/2} \ ,
\label{dpe} \eeq
which is similar to the confluent hypergeometric function $\Psi$.
These solutions can only be obtained by applying $\pi$ or $\pi^{\dagger}$ to a
neighbor polynomial solution. In the appendix, we report a table of the
polynomial eigen-functions (\ref{pol}),(\ref{pol2} ) for the first few
values of $k$ and $\ell$, and give examples of the action of $\pi$.
Note that no normalizable solutions are found with energy
$k = 0$; thus, there is the double of the Landau-level gap between the
lowest available level $( k = - 1)$ and all the higher ones
$( k = 1, 2, ... )$.
We do not have a physical interpretation of this result.

In conclusion, the complete energy spectrum is discrete and
given by eq. (\ref{spe}) for $k=-1,1,$ $2,3,...$, with each level
infinitely degenerate in angular momentum ($\ell \ge -k$).

Let us discuss more precisely the action of the magnetic translation operators
$\pi$ and $\pi^{\dagger}$ and show that it closes on these solutions,
thus ensuring their translation (and gauge) invariance.
Equations (\ref{alg},\ref{idd}) imply $\pi^{\dagger}\pi\psi_{k,-k}=0$ and
$\pi\pi^{\dagger}\psi_{k,-k-1}=0$: by explicit calculation
(see the appendix), we actually find
$\pi\psi_{k,-k}=0$, leading to the following pattern,
\beq
\begin{array}{rclll}
\cdots\longrightarrow & \psi_{k,-k-1}&
{\buildrel \pi^{\dagger}\over\longrightarrow} &
\psi_{k,-k}\longrightarrow &\psi_{k,-k+1} \longrightarrow \cdots \\
& 0 & {\buildrel \pi\over\longleftarrow} &
\psi_{k,-k}\longleftarrow &\psi_{k,-k+1} \longleftarrow \cdots
\end{array}
\label{sep}\eeq
Therefore, the representation of the $(\pi,\pi^\dagger)$ gauge algebra is
not {\it fully decomposable} into normalizable $(\ell\ge -k)$ and
non-normalizable $(\ell<-k)$ states:
the former are mapped by $\pi$ and $\pi^{\dagger}$ into themselves,
while the latter are also mapped by $\pi^{\dagger}$ into normalizable ones.
Nevertheless, the projection of the non-normalizable states to zero is
consistent\footnote{
Note that our analysis does not exclude the existence of other solutions
for some $\ell \neq 0 $ isolated values. However, these would not be acceptable
because they would not close under the action of $\pi$.}.
Actually, the same non-decomposable representation occurs in the elementary
Landau levels, because the action of the operators $\pi^L$ and
$(\pi^L)^{\dagger}$ on the Landau wave-functions is the same as (\ref{sep}).
Therefore, non-decomposability seems to be a rather general property of
gauge invariance in the Hamiltonian formalism.

We now discuss the completeness of the basis of eigen-functions we have found.
For $\ell = 0$, the eigen-functions (\ref{hyp}) are linear combinations of the
polynomials
$ \{ r^{2n} e^{-r^2/2}, n > 0 \} $, which form a complete basis for
square integrable functions of $ ( r^2 )$ with vanishing
zero mode. For $\ell \neq 0$, the space of solutions cannot be easily
defined in mathematical terms, due to the involved $r=0$ boundary conditions,
which can be either $\psi \simeq e^{i \ell \theta} r^0$ or
$\psi \simeq e^{i \ell \theta} r^{| \ell |}$.
Therefore, the issue of completeness cannot be easily addressed
for the $\ell \neq 0$ subspaces, which are, nevertheless, isomorphic
to the $\ell=0$ complete basis by the $\pi$ action. Note that the singular
behavior $\psi \simeq e^{i \ell \theta} r^0 $ at $ r = 0 $, is fully
acceptable, because it cancels in gauge invariant quantities like
$\rho^a\rho^a \propto \psi^{\dagger} \psi $.

\section{Spontaneous symmetry breaking and the excitations of the anyon fluid}

We have been describing the Abelian and non-Abelian $ U(2)$
anyon fluids, which are non-relativistic gauge theories of the Chern-Simons
type, and we have shown the spontaneous breaking of the corresponding global
symmetries U(1) and $ U(2) \rightarrow U(1) $, respectively.
It is interesting to discuss the analogies and differences with
the four-dimensional Yang-Mills theory of the Standard Model of
electroweak interactions, and identify the excitations of the anyon fluid
with the non-relativistic analogues
of Goldstone and Higgs particles, if possible.
There are two basic differences:

i) the non-relativistic matter fields have half of the degrees of freedom of
their relativistic counterparts, because the latter describe both particles
and antiparticles;

ii) the Chern-Simons gauge field has no propagating physical degrees of
freedom and thus cannot lead to the Higgs phenomenon.

Let us first recall the superfluid, which is the canonical example
of spontaneous symmetry breaking in non-relativistic field theory:
\beq
H = \int d^2 {\bf x} \left( \frac{1}{2m} {| \partial_i \Psi |}^2 + {g\over 2}
{| \Psi |}^4 \right) \ ,\qquad g > 0 \ .
\label{sft} \eeq
Due to Bose condensation, the field acquires the ground-state expectation
value $\langle \Psi\rangle =\sqrt{ \rho_0 }$, which breaks the $U(1)$ global
symmetry of particle number conservation.
Small excitations around the mean field are diagonalized by the
Bogoliubov transformation. This leads to a massless excitations with sound
velocity $ v_s \propto \sqrt{ g \rho_0 } $, controlled by the repulsive
interaction; moreover, the ground state does not have a definite particle
number, due to the Bogoliubov rotation.

The usual description \cite{flu} in terms of radial $(\hat{\rho})$ and
phase $(\theta)$ components of the field,
$\Psi = \sqrt{ \rho_0 + \hat{\rho} } \ e^{i\theta}$,
can be easily compared to the current algebra description (\ref{cr}).
The current is represented as,
\beq
J_i = \frac{1}{m}\ {\rm Im}\ \left( \psi^{\dagger} \partial_i \psi \right) =
\frac{\rho_0 + \hat{\rho}}{m}\ \partial_i \theta \ ,
\label{rph} \eeq
and the approximate commutation relations (\ref{app}), (\ref{app2}) become,
\beq
\left[ J_i , \hat{\rho} \right] \simeq
\frac{i\rho_0}{m} \ \partial_i \delta ( {\bf x}- {\bf y} )\
\longrightarrow \left[ \theta, \hat{\rho} \right] =
i \delta ( {\bf x} - {\bf y} ) \ .
\label{acr} \eeq
Therefore, the would-be relativistic Higgs ($\hat{\rho}$) and Goldstone
($\theta$) fields are conjugate variables in the non-relativistic theory,
the superfluid massless mode is a Goldstone particle and there is no
non-relativistic analogue of the Higgs particle.

The Abelian anyon fluid is very similar to the superfluid. The
``microscopic'' mechanism leading to $\langle\rho\rangle=\rho_0$
is not the Bose condensation - there is no macroscopic occupation of a
single energy level, rather a macroscopic number of particles at the
same energy.
Nevertheless, there is spontaneous breaking of the global $U(1)$ symmetry,
because the Bogoliubov transformed ground state has no definite particle
number (see sect 2.3).

One can find a closer relation with the usual superfluid by formally
integrating out the Chern-Simons field.
The resulting self-interacting  matter theory can possibly reduce to
(\ref{sft}) for small fluctuations around the saddle-point approximation.
In this sense, we can consider the anyon superfluid as a non-relativistic
example of {\it dynamical} symmetry breaking \cite{iz}.
However, this is a peculiar example, where the self-interaction should have
special normal-ordering effects, as in sec. 2.3, 2.4, which determine
different effective $|\Psi|^4$ interactions for boson-based $(g\propto 1/\k)$
and fermion-based $(g\propto 1)$ anyons.

The current algebra approach is general enough to handle this non-standard
mechanism of spontaneous symmetry breaking.
By expanding again $\Psi$ into $(\rho,\theta)$ components, one obtains
the current $J_i\simeq\rho_0\left( \partial_i\theta-A_i\right)/m$;
this is still
purely longitudinal, because the Chern-Simons gauge field can be (locally)
reduced to a pure gauge, $A_i=\partial_i\alpha$,
\beq
J_i \simeq {\rho_0 \over m}\ \partial_i (\theta-\alpha ) \ .
\eeq

Therefore, the current algebra (\ref{app}), (\ref{app2}) is the same as in the
superfluid (\ref{acr}) and the anyon massless mode is a Goldstone particle.
Note that the current, i.e. $ ( A_i - \partial_i \theta )$,
is the fundamental quantity of the Landau-Ginzburg theory of superconductivity
\cite{wein}, but it has there a completely different meaning, because the
gauge field has transverse degrees of freedom.

The $U(2)$ non-Abelian anyon fluid can be described in similar terms.
The mean field value $ \langle \rho^a\rangle=\delta^a_3 \ \rho_0/2$
breaks the $SU(2)$ global symmetry to the $U(1)$ subgroup generated by
$ ( 1 - \sigma_3 ) / 2 $.
Although $\Psi_r$ has vanishing ground-state value, it is still convenient
to introduce the parametrization of the Standard Model \cite{iz},
\beq
\Psi_r = \exp\left(i \frac{\theta^a \sigma_a}{2}\right)\left( \begin{array}{c}
\sqrt{ \rho_0 + \hat{\rho} } \\ 0  \end{array} \right) \ ,
\label{ghi}\eeq
where $\theta^a$ are the three would-be relativistic Goldstone particles and
$\hat\rho$ the would-be Higgs one.
The non-relativistic field $\Psi_r$ describes only two of these
degrees of freedom, while the other two are conjugate momenta.
The approximate current algebra (\ref{acNA},(\ref{acNA2}) can be
rewritten as follows:
\barr
J_i \simeq {\rho_0\over m}\ \partial_i \theta^3\ , \ \ \
\rho^3 \simeq \rho_0 + \hat{\rho} & \longrightarrow  &
\left[ \theta^3({\bf x}), \hat{\rho} ({\bf y}) \right]
\simeq i \delta ( {\bf x } - {\bf y } ) \ ,
\nonumber\\
\rho^1 \simeq \rho_0\ \theta^2\ \ \ \ , \ \ \
\rho^2\simeq - \rho_0 \theta^1 & \longrightarrow &
\left[ \theta^1({\bf x}), \theta^2 ({\bf y}) \right]
\simeq \frac{1}{\rho_0} \delta ( {\bf x } - {\bf y }  ) \ .
\earr
The density $\hat{\rho}$ and the isospin-diagonal phase $\theta^3$ represent
the Goldstone particle as in the Abelian case; this excitation does not have a
well-defined isospin number.
The other pair of would-be Goldstones are conjugate variables; they
do not undergo the Bogoliubov transformation, because the isospin-down
number is conserved by the remaining $U(1)$ symmetry.
The effective repulsive interaction
$ \left( 4 /|\kappa| - 1 /\kt \right) {( \rho^a )}^2 $
induced by the Chern-Simons field gives a
mass gap to this excitation.

\section{Concluding remarks}

In this paper, we analyzed the $U(2)$ Chern-Simons theory coupled to
non-relativistic matter with isospin $1/2$.
We applied the mean field approximation developed in the refs.
\cite{fhl},\cite{carlo}, and uncover a phase of the theory where
the global $U(2)$ symmetry breaks spontaneously to the $U(1)$ one.
Besides one Goldstone excitation already present in the Abelian model,
we found a massive excitation with non-trivial non-Abelian dynamics.
Therefore, the phenomenon of superfluidity and superconductivity, originally
discovered by Laughlin \cite{fhl}, extends smoothly into
this phase of the non-Abelian anyon fluid.

This theory can also be consider as a toy model of the Standard Model of the
electroweak interactions: in section five, we clarified the analogies
and differences between the Chern-Simons theory and the four-dimensional
Yang-Mills theory.
Although there are many simplifying features, the non-Abelian anyon
fluid is an instructive example where a non-Abelian gauge theory can be
explicitly solved in the low-energy limit.

Another interesting aspect of this theory
is its close relation with gravity in $( 2 + 1 )$ dimension.
Actually, the Chern-Simons action for the non-Abelian group $ISO(2,1)$
( respectively $SO(4)$ ) can be rewritten as the Einstein-Hilbert action
(respectively with cosmological constant) \cite{grav}.
A suitable coupling of matter fields to gravity might lead to a theory
similar to the non-Abelian anyon fluid:
the mean-field approximation might describe a phase of semiclassical
cosmology, where the metric is ``induced'' by the
(dynamical) symmetry breaking \cite{cap}.

{\bf Acknowledgments}

We would like to thank C.A. Trugenberger for many useful discussions
and a careful reading of the manuscript.
We also acknowledge the discussions with S. Catani, P. Menotti and
M. Mintchev. This work was supported in part by the CERN Theory Division
and the EEC Program ``Humal Capital and Mobility''.

\bigskip
\appendix
\section{ Details of the eigenvalue problem}

{\bf Examples of $\ell\neq 0$ eigen-functions}

The polynomial solutions (\ref{pol}), denoted by the
points $(\bullet)$ in fig. 2, are of the form,
\beq
\psi={\rm e}^{il\theta}\ (2+r\partial_r)\ \varphi\ ,\quad
\varphi (r)=r^\ell\ \left( a_0 +a_1r^2+\cdots +a_{k-1}\ r^{2k-2}\right)\
{\rm e}^{-r^2/2}\ ,
\eeq
and exist for $k\ge 2$ , even, and $\ell\ge 0$.
The simplest ones are listed hereafter $(\ell_n\equiv(\ell+n))$:
\beq
\begin{array}{l|llllll}
k & a_0 (\ell)& a_1        & a_2          & a_3         & a_4       & a_5 \\
\hline
2 & -\ell_1 & 1          &              &             &           &     \\
4 & -\ell_4\ell_3\ell_1 & \ell_3(3\ell+8)& -3\ell_3& 1 &          &     \\
6 & -\ell_6\ell_5\ell_4\ell_3\ell_1
  & \ell_6\ell_5\ell_3(5\ell+12)& -2\ell_5\ell_3(5\ell+26)
  & 2\ell_5(5\ell+22) & -5\ell_5          & 1
\end{array}\eeq
The other type of polynomial solutions, denoted by $(\circ)$ in fig. 2,
are of the form,
\beq
\psi={\rm e}^{il\theta}\ (2+r\partial_r)\ \varphi\ ,\quad
\varphi (r)= \left( {b_{-1}\over r^2} +b_0+ b_1r^2
+\cdots +b_{k-1+\ell/2}\ r^{2k+\ell-2}\right)\
{\rm e}^{-r^2/2}\ ,
\eeq
and exist for $k\ge -1$ and $\ell\ge -k$, $\ell$ even. The simplest ones
are listed hereafter:
\beq
\begin{array}{l|l|llll}
k  & \ell     & b_{-1}    & b_0          & b_1         & b_2   \\ \hline
-1 & 2        & 1          &              &             &       \\
-1 & 4        & 4          & 1            &             &        \\
-1 & 6        & 24         & 8            & 1           &        \\
\ 1 & 2        & -2         & 0            & 1           &        \\
\ 1 & 4        & -16        & -4           & 0           & 1      \\
\end{array}\eeq
Note that for some even, negative, values of $\ell$, these two
types of solutions actually merge.

\bigskip

\noindent{\bf Examples of the action of $\pi$ and $\pi^{\dagger}$}

Let us derive some of the non-polynomial solutions, denoted by
$(\times)$ in fig. 2, by applying $\pi$ to a polynomial solution
with neighbor value of $\ell$. Consider for example the lower energy level
$k=-1$, and find $\psi_{-1,1}$:
\barr
\psi_{-1,1} \propto \pi\psi_{-1,2}
&=& \left(\partial_z +{\bar{z}\over 2}-\bar{z}
{1\over (1+r\partial_r)(2+r\partial_r)} \right)\
\left( 2+r\partial_r \right)\ {{\rm e}^{i2\theta}\over r^2}\ {\rm e}^{-r^2/2}
\nonumber\\
&=& -{\rm e}^{-r^2/2}\ \partial_z {\rm e}^{i2\theta}\ -\
{\rm e}^{i\theta}\ {1\over r\partial_r}\ {1\over r}\ {\rm e}^{-r^2/2}
\nonumber\\
&=& {\rm e}^{i\theta}{1\over r\partial_r}\ r\ {\rm e}^{-r^2/2} =
-\ {\rm e}^{i\theta}\ {r\over 2}\ \Psi\left(1,{3\over 2},{r^2\over 2}\right)\ .
\earr
In this derivation, we used some formal properties of $(r\partial_r)^{-1}$
which follows by integration by parts of (\ref{iac}),
and, at the very end, its explicit form (\ref{asy}).
The further application of $\pi$ yields:
\beq
\pi\psi_{-1,1}=\left(\partial_z +{\bar{z}\over 2}-
\bar{z}{1\over (1+r\partial_r)(2+r\partial_r)} \right) {z\over r}\
{1\over r\partial_r}\ r\ {\rm e}^{-r^2/2}= \ 0\ .
\label{chk}\eeq
Actually, this vanishing result can be found more easily
by collecting the common denominator $1/(1+r\partial_r)$.
Equation (\ref{chk}) verifies the closure of the gauge algebra on
the physical solutions with energy $k=-1$, as indicated in the
diagram (\ref{sep}).
Another non-trivial action in this diagram is given by $\pi^{\dagger}$
applied to the unphysical logarithmic solution $\psi_{-1,0}$  found
in (\ref{hyp}), which has the explicit form,
\beq
\psi^{(\rm log)}_{-1,0}={2+r\partial_r\over r\partial_r}\
{1\over r^2}\ {\rm e}^{-r^2/2} = -{1\over r\partial_r} \ {\rm e}^{-r^2/2}\ .
\eeq
The action of $\pi^\dagger$ is computed as follows,
\beq
\pi^{\dagger}\psi^{(\rm log)}_{-1,0}={z\over 2}\left(
{1\over 2+r\partial_r} + {1\over r\partial_r}-
{2\over (1+r\partial_r)( 2+r\partial_r)}\right) {\rm e}^{-r^2/2}=
\psi_{-1,1}\ .
\eeq

One can similarly compute that the $(\ell=0)$ logarithmic solutions
$\Psi$ in (\ref{hyp}), for any  value of $k\neq 1,2,3\dots$, are mapped
by $\pi$ and $\pi^{\dagger}$ into $\ell=\pm 1$ solutions with
non-integrable behavior $\psi\simeq r^{-|\ell|}$; actually,
it is sufficient to use the $(r\to 0)$ expansion of these eigen-functions
\cite{bat} which is reported in (\ref{logg}).

\bigskip
\noindent{\bf Solutions with $r^{-4}$ asymptotics at $(r\to\infty)$}

The study of the asymptotic behaviors shows that these solutions
are also solutions of the inhomogeneous reduced problem (\ref{soe}),
$[(r\partial_r)^2 -r^4 +2\lambda r^2 -4]\phi^{(-4)}=1$.
It is sufficient to solve it in the case $k\neq 1,2,3\dots$.
The inhomogeneous solutions can be obtained from the Green function,
\beq
G(r,\rho)={u_1(r)u_2(\rho) \Theta(\rho -r) +u_1(\rho)u_2(r)\Theta(r-\rho)
\over a_2(\rho) W(\rho)} \ .
\eeq
In this equation, $u_1$ and $u_2$ are the two independent,
homogeneous solutions vanishing
at $(r\to 0)$ and $(r\to\infty)$, which are given by
the $\Phi$ and $\Psi$ confluent Hypergeometric functions in (\ref{hyp}),
respectively.
Moreover, $a_2 $ is the coefficient of second-order term in the differential
equation and $W$ is the wronskian, $a_2(\rho)W(\rho)\propto \rho$.
The resulting expression for the Green function integrated against
the source can be expanded for asymptotic values of $r$.
For $(r\to\infty)$, one recover the $r^{-4}$ behavior by the
cancellation of the positive and negative exponentials of
$\Phi$ and $\Psi$. For $(r\to 0)$, one finds the behavior,
\beq
\phi^{(-4)}_{k,0}\propto {1\over \Gamma(1-k)}
\left[ 1 + O\left(r^2\right) +O\left(r^2\log r^2\right) \right]
{\rm e}^{-r^2/2} \ ,
\eeq
which leads to a logarithmic behavior for $\psi^{(-4)}_{k,0}$ and
a non-normalizable one for $\psi^{(-4)}_{k,1}=\pi^\dagger \psi^{(-4)}_{k,0}$,
by the same mechanism discussed above.

\def\NP{{\it Nucl. Phys.\ }}
\def\PRL{{\it Phys. Rev. Lett.\ }}
\def\PL{{\it Phys. Lett.\ }}
\def\PR{{\it Phys. Rev.\ }}
\def\IJMP{{\it Int. J. Mod. Phys.\ }}
\def\MPL{{\it Mod. Phys. Lett.\ }}

\end{document}